\documentclass[aps,prb,amsmath,amssymb,superscriptaddress,twocolumn]{revtex4-2}
\usepackage{CJK}
\usepackage[utf8]{inputenc}
\usepackage{graphicx}
\usepackage{dcolumn}
\usepackage{amsmath}
\usepackage{bm}
\usepackage{physics}
\usepackage{verbatim}
\usepackage{color}
\usepackage{tabularx}
\usepackage{xcolor}
\usepackage{cancel}
\usepackage{hyperref}
\usepackage{xr}
\definecolor{gold}{rgb}{0.83, 0.69, 0.22}

\newcommand\COMMENTED[1] {}

\newcommand{\sys}{2D HEG}
\newcommand{\up}{\uparrow}
\newcommand{\dn}{\downarrow}

\begin{document}

\title{The two-dimensional homogeneous electron gas 
with symmetric dual-gate screening: exchange-correlation functional and other ground-state properties}

\author{Yiqi Yang} 
\email{yyang25@wm.edu}
\affiliation{Department  of  Physics,  College  of  William  \&  Mary,  Williamsburg,  Virginia  23187, USA}
\author{Yubo Yang}
\affiliation{Center for Computational Quantum Physics, Flatiron Institute, 162 5th Avenue, New York, New York, USA}
\author{Kun Chen}
\affiliation{CAS Key Laboratory of Theoretical Physics, Institute of Theoretical Physics, Chinese Academy of Sciences, Beijing 100190, China}
\author{Miguel A. Morales}
\affiliation{Center for Computational Quantum Physics, Flatiron Institute, 162 5th Avenue, New York, New York, USA}
\author{Shiwei Zhang}
\affiliation{Center for Computational Quantum Physics, Flatiron Institute, 162 5th Avenue, New York, New York, USA}

\begin{abstract}
The two-dimensional (2D) homogeneous electron gas (HEG) is a fundamental model in quantum many-body physics. It is important to theoretical and computational studies, where exchange-correlation energies computed in it serve as the foundation for density-functional calculations. It is also of direct relevance to a variety of experimental settings, especially with the rapid recent growth in 2D materials and moir\'e systems. In these experiments, metallic gates are often present, which screen the Coulomb interaction between electrons. The effect of the screening can qualitatively change the behavior of the 2D HEG, and requires accurate many-body computations to capture. In this work, we perform state-of-the-art diffusion Monte Carlo (DMC) calculations in the 2D HEG subjected to symmetric dual-gate screening. We systematically compute the correlation energy across a range of densities and gate separations for both spin unpolarized and fully polarized systems. A global fit is obtained for the correlation energy, using these data and imposing various limiting behaviors obtained from perturbation analysis. The new functional will allow density-functional calculations to be performed for a variety of realistic experimental setups which can accurately account for the presence of gates. We also investigate how the gate screening affects the bulk modulus, pair correlation function, and the structure factor of the 2D HEG, which can potentially be probed in experiments.
\end{abstract}
\maketitle

\section{Introduction}

There has been an explosion of research activities in two-dimensional (2D) electronic systems, driven by rapid developments in moir\'e materials. These systems exhibit enhanced correlations compared to their bulk counterparts due to the suppressed kinetic energy and reduced screening effect~\cite{Ashcroft_Book1976}. These 2D materials offer greater flexibility in manipulation through stretching, doping, and the application of external fields. Recent advances in stacking and twisting 2D systems have demonstrated their potential as quantum simulators for paradigm models~\cite{Ando_RMP1982,Tang_NATURE2020,Wu_PRL2018} and as platforms for studying correlation-driven phases, including unconventional superconductivity~\cite{Cao_NATURE2018,Xia_arxiv2024,Guo_arxiv2024} and the fractional quantum anomalous Hall effect~\cite{Park_NATURE2023,Lu_NATURE2024}. 

In experimental setups, metallic gates are often employed to apply external electric fields and adjust the charge density of 2D systems. The gates screen the Coulomb interaction, which can quantitatively or qualitatively affect the physics of the 2D system of interest. It is worth noting that this screening effect is not just a necessary inconvenience, but is sometimes an advantage, as it can be leveraged experimentally to control the strength and range of the Coulomb interactions. By varying gate distances, researchers have explored the competition between magnetic orders~\cite{Tang_NN2023,Wu_PRL2018,Stepanov_Nature2020} and investigated transport properties~\cite{Nuckolls_NRM2024,Kim_NC2020,Saito_NP2020,Stepanov_Nature2020}. Additionally, the metallic gates have been shown to stabilize superconductivity~\cite{Liu_Science2021}.

To better understand or model experiments, the presence of the metallic gates needs to be properly addressed. This is challenging because the modification to the interaction can lead to significant effects that are non-perturbative and require accurate computational treatment. (For example, how does the presence of the gates affect the existence of the Wigner crystal phase as the location of the transition?) The most widely applied computational approach is perhaps density functional theory (DFT) calculations under the local density approximation (LDA). These calculations can provide quantitative descriptions of atomic-level details and are often used in conjunction with experiments for parameter screening, and theoretical analysis. As an independent-electron approach, DFT/LDA is expected to break down in many situations of strong electron correlation, but it is often surprisingly accurate. Even in 2D moir\'e systems, it was shown to produce reasonable results for several strongly correlated phases~\cite{Yang_PRL2024,Yang_arxiv2024}. 
Given its wide accessibility, we can expect that DFT approaches will continue to serve as a first-stop computational tool in the study of 2D materials. 
However, the current DFT exchange-correlation functionals are derived from the 2D HEG with bare Coulomb interaction~\cite{Kwon_PRB1993,Attaccalite_PRL2002,Drummond_PRL2009}, without any consideration of screening from the gates. This is a major deficiency that can significantly impact the applicability of DFT in modeling moir\'e systems.  

In this work, we address this problem, by performing state-of-the-art diffusion Monte Carlo (DMC)~\cite{Kalos_BOOK2008,Hammond_BOOK1994,Foulkes_RMP2001} calculations on the \sys\ subjected to symmetric-gate screening.
We systematically study the ground state across a range of densities and screening strengths. At each condition, we compute the total energy for both spin-polarized and spin-unpolarized systems and subtract the corresponding Hartree Fock energy to obtain correlation energy. Asymptotic behaviors in the strongly screened limit are derived from perturbation analysis, as well as the leading order corrections to the usual 2D HEG (bare Coulomb interaction) energy in the weakly screened limit. 
The computed correlation energies are then fitted to a global form for all $(r_s,d)$ values for the unpolarized and fully polarized systems, respectively, with the form satisfying the limiting conditions mentioned above. A functional for partial polarization is then obtained using a commonly adopted formula~\cite{Koskinen_PRL1997,Gori-Giorgi_IJQC2003} to interpolate the spin-polarized case and spin-unpolarized case.  The fitted correlation energy can serve as a local spin density approximation (LSDA) functional, enabling efficient and more accurate DFT calculations to model 2D systems in the presence of metallic gates.

In addition to obtaining the exchange-correlation functional, we also obtain several ground-state properties of 
the \sys~under dual-gate screening, including the spin-resolved pair correlation function $g(r)$, the static structure factor $S(k)$, and the bulk modulus $B$. Besides their intrinsic value in providing insights into the physics of the 2D HEG as a function of the screening strength on the Coulomb interaction, some of them can be directly accessed by experiment. For instance, in the absence of gate screening, $B$ changes sign at a critical electron density, a phenomenon detected in experiments~\cite{Eisenstein_PRL1992}.

The rest of the paper is organized as follows. We specify the model in detail in Sec.~\ref{sec:model}, and describe our computational method in Sec.~\ref{sec:method}. In Sec.~\ref{sec:MF}  we discuss the Hartree-Fock results, which are used as a reference for defining the correlation energy. The perturbative analysis used to obtain the asymptotic limits is then presented in Sec.~\ref{sec:Analytic_results}. We give our global fitting form and the final result of the exchange-correlation functional in Sec.~\ref{sec:Global_fit}. Additional properties of the screened 2D HEG, including the systematic dependence of pair correlation functions on the screening strength, are discussed in Sec.~\ref{sec:Other_properties}, before we conclude in Sec.~\ref{sec:Conclusions}.

\section{Model}
\label{sec:model}

The \sys~model
describes electrons confined to a two-dimensional plane in the presence of a uniform compensating charge background.
When metallic gate electrodes are present the Coulomb interaction between electrons are screened resulting in the Hamiltonian:
\begin{equation}
\label{eq:H}
\hat{H}=-\frac{1}{2}\sum_i\nabla_i^2+\frac{1}{2}\sum_{i\neq j}v\left(\abs{\mathbf{r}_i-\mathbf{r}_j},d\right) +{\rm b.g.}\,.
\end{equation}
Hartree atomic units are used throughout the paper, where length is measured in units of the Bohr radius $a_B$ and energy is measured in Hartree. We consider $N$ electrons in a periodic simulation cell, a square of side length $L$.
In this work, calculations are performed with fixed spin polarization $p=(N_{\up}-N_{\dn})/(N_{\up}+N_{\dn})$, where $N_\up$ and $N_\dn$ are the number of spin-up and spin-down electrons, respectively
($N_\up+N_\dn=N$). The physics of the system is then controlled by two
parameters: Wigner-Seitz radius $r_s$ ($N \pi r_s^2=(L/a_B)^2$), and the distance from the \sys~to the metallic gates $d a_B$.
Our calculations will focus on $r_s \in [0,30]$, i.e., above the Wigner crystal transition density in the unscreened 2D HEG.
In the remainder of the paper, we use two dimensionless parameters to specify the physical system: $r_s$ and the strength of screening $\mu\equiv r_s/d$. The bare Coulomb (unscreened) limit is obtained at $d\rightarrow\infty$ ($\mu \rightarrow 0$), while the non-interacting limit is approached as $\mu \rightarrow\infty$.

Given two metallic gates parallel and equal distance to the \sys, the screened interaction $v(r,d)$ can be evaluated by the method of images in electrostatics,
\begin{equation} \label{eq:vimgs}
v(r,d) = \sum_{m=-\infty}^{\infty} (-1)^m/\sqrt{r^2+(2md)^2}.
\end{equation}
The sum in Eq.~(\ref{eq:vimgs}) is conditionally convergent and should be evaluated with proper regularization techniques (see details in Appendix~\ref{appendix:gateScreening}). 
Written in Fourier space, this potential has a simple closed form: 
 \begin{equation}\label{eq:kspace}
     v(q,d)=\frac{2\pi}{q}~\mathrm{tanh}(qd).
 \end{equation}
Asymptotically, the screened interaction decays exponentially in real space $v\left(r\gg d\right)\simeq e^{-\frac{\pi r}{2d}}/\sqrt{r}$~\cite{Zangwill_book2013,Valenti_PRL2024}.

\section{Methods}
\label{sec:method}

We employ state-of-the-art fixed-node diffusion Monte Carlo method (FN-DMC)~\cite{Ceperley_PRB1978,Ceperley_PRL1980,Attaccalite_PRL2002,Drummond_PRL2009} to study the Hamiltonian of Eq.~(\ref{eq:H}), as defined in Sec.~\ref{sec:model}. The implementation of FN-DMC in QMCPACK~\cite{Kim_JPC2018,Kent_JCP2020} is used with appropriate modifications to treat the screened Coulomb interaction.
To compute the interaction potential $v(r,d)$, the Ewald method is used to evaluate Eq.~(\ref{eq:kspace}) for weak to modest screening ($\mu\le 1$). When the screening is strong ($\mu>1$), we directly sum the screened Coulomb potentials in the real space Eq.~(\ref{eq:vimgs}).

The ground-state energy {\it per electron\/}, $\epsilon$,  is computed as a function of $r_s$, $\mu$, and $p=0,1$.
This gives the correlation energy, defined as the difference between the many-body ground-state energy and the mean-field (restricted Hartree-Fock) counterpart: $\epsilon_c\equiv \epsilon-\epsilon_{\text{HF}}$. We then fit $\epsilon_c(r_s,\mu,p)$ subject to all the analytic constraints from the limiting behaviors derived from perturbative analysis. Our QMC calculations are mostly performed for the unpolarized ($p=0$) and fully polarized ($p=1$) systems. 
We interpolate the results to intermediate polarizations, with explicit QMC calculations performed at select $p$ values to validate the quality of the interpolation.

As mentioned, we focus on the density range of $r_s$ between $0$ and $30$, which is in the liquid phase in the unscreened 2D HEG. 
The trial wave function (TWF) is chosen to be the Slater determinant of plane wave orbitals multiplied by a two-body B-spline Jastrow function. A backflow transformation using two-body B-spline is included in the Slater determinant to further improve the form of TWF.
The TWF is optimized by variational Monte Carlo (VMC) ~\cite{Kim_JPC2018} before being used in FN-DMC. 

Separate FN-DMC calculations are carried out with different values of the time step, and a linear extrapolation is performed to eliminate the time step error. Autocorrelation is removed by the reblocking (rebinning) procedure~\cite{Shi_PRE2016}. The walker size in all the calculations is set to be $2304$, which is large enough for the bias from population control to be negligible within our statistical errors. 

QMC calculations are performed in finite-size simulation cells with a finite number of electrons, $N$. Our FN-DMC calculations use $N=42$, $90$, and $122$ for both the spin-polarized case and the spin-unpolarized case. The ground-state energy and properties computed from these have systematic errors with respect to the desired values in the thermodynamic limit. We take several measures to remove these finite-size effects.
First, we apply twisted boundary conditions (TBC) in our calculations~\cite{Ceperley_PRE2001}. The twist angles are randomly sampled, and the number of TBC $N_k$ is set to around $10^3/N$. Second, we compute $\epsilon_c$ with correlated twists, 
averaging the differences~\cite{Kwee_thesis2008} between the total energy and the HF energy over the same set of twists.
We then apply an extrapolation to the correlation energy: $\epsilon_c^{\infty}=\epsilon_c^N+a/N+b/N^2$ (see Appendix~\ref{appendix:FS}), where $\epsilon_c^N$ is the twist-averaged correlation energy per electron for system with $N$ electrons, and $\epsilon_c^{\infty}$ the correlation energy per electron in thermodynamic limit, which is obtained via a fit 
together with the other two parameters $a$ and $b$. 
We find that the finite-size correction given by a lower level of theory, i.e., FN-DMC with a Slater-Jastrow TWF (i.e., omitting backflow) is consistent with that obtained using backflow in systems we have tested. Therefore, we estimate finite-size corrections for the remaining systems without backflow, to reach a larger system size $N$ with less computational cost.
All the pair correlation functions are obtained from systems of $42$ electrons, using Slater-Jastrow-Backflow TWF. They are averaged over $25$ twist angles.

\section{The Hartree-Fock
energy}\label{sec:MF}

The  exchange energy is defined as~\cite{Pines_book1999},
\begin{equation}\label{eq:ex_convention}
\epsilon_x=\bra{\psi_0}\hat{H}_{\mathrm{int}}\ket{\psi_0}/N,
\end{equation}
where $\ket{\psi_0}$ is the non-interacting ground state. In the absence of metallic gates, the HF energy per electron for 
\sys~is known~\cite{Giuliani_Book2005}: 
\begin{equation}\label{eq:bc_HF}
\epsilon_{\mathrm{HF}}^{\mathrm{Coul}}=\frac{1}{2r_s^2}\frac{P_+^4+P_-^4}{2}-\frac{4\sqrt{2}}{3\pi}\frac{1}{r_s}\frac{P_+^3+P_-^3}{2},
\end{equation}
where $P_{+(-)}=\sqrt{1+(-)p}$.
The first term is the kinetic energy and the second term is the exchange energy $\epsilon_x^{\mathrm{Coul}}(r_s,p)$.
We find that the effect of gate screening on the exchange energy can be described by 
\begin{equation}
\epsilon_x(r_s,\mu,p)=\epsilon_x^{\mathrm{Coul}}(r_s,p=0)F(\mu,p).
\end{equation}
where $F(\mu,p)$ is given in Eq.~(\ref{eq:F}). For convenient implementation in computer routines, we provide a fit to the exchange energy in Appendix~\ref{appendix:HF}.

\section{Analytic constraints and asymptotic behaviors}\label{sec:Analytic_results}

The correlation energy in the absence of gate screening was evaluated by QMC~\cite{Ceperley_PRB1978,Kwon_PRB1993,Attaccalite_PRL2002,Drummond_PRL2009} and perturbation analysis~\cite{Rajagopal_PRB1977,Seidl_PRB2004}.
In a previous study, Attaccalite et al.~\cite{Attaccalite_PRL2002} provided an analytical fit to $\epsilon_c^{\mathrm{Coul}}$ in a Pad\'e form. 
Zecca et.~al.~\cite{Zecca_PRB2004} studied the 3D HEG with screened interaction (in the form of a complementary error function). 
We use a functional form for the correlation energy of the screened 2D HEG which follows the ideas of these previous studies. It has a  Pad\'e form guided by asymptotic behaviors given by perturbation analysis.
We fit the FN-DMC results in the spin-unpolarized and the spin-polarized cases separately~\cite{Vosko_CJP_1980,Perdew_PRB1992,Attaccalite_PRL2002,Zecca_PRB2004}. The correlation energy at intermediate spin polarization is obtained by interpolation, validated by spot checks with FN-DMC calculations at finite $p$. We also fit our correlation energy in the absence of gate screening, using Attaccalite et al.'s form in the spin-unpolarized limit and spin-polarized limit, as is described in Appendix~\ref{appendix:EcBC}.

The actual forms are presented in Sec.~\ref{sec:Global_fit}, together with the parameter values from the fit. 
Below we describe the analysis which provides the limiting behaviors imposed on the functional form.

\subsection{Strongly screened limit}
When the electron-electron interaction is strongly screened ($\mu \gg 1$), the system is a fluid with short-range interactions.
The correlation energy can be calculated by perturbation theory.
In the spin-unpolarized case, 
\begin{equation}\label{eq:pert_up}
\epsilon_c(r_s\geq0, \mu, p=0)\xrightarrow{\mu\rightarrow\infty}-0.5\ln(\mu)/\mu^2.
\end{equation}

In the spin-polarized case,
\begin{equation}\label{eq:pert_p1}
\epsilon_c(r_s>0, \mu, p=1)\xrightarrow{\mu\rightarrow\infty} r_s/\mu^3
\end{equation}
for any finite density $r_s>0$. 

If we first take the limit of $r_s\rightarrow0$ while keeping $\mu$ a large parameter, we obtain the following behavior:
\begin{equation}\label{eq:pert_p2}
 \lim_{r_s \rightarrow 0} \epsilon_c(r_s, \mu, p=1) \xrightarrow{\mu \rightarrow \infty} 1/\mu^4
\end{equation}

The derivations can be found in Appendix~\ref{appendix:perturbation}.

\subsection{High-density limit}

In the high-density limit, the system is weakly interacting.
The leading-order contribution to the correlation energy comes from the second-order perturbation, as discussed in Appendix~\ref{appendix:perturbation}. 
In the infinite density limit $r_s=0$, perturbations beyond second order vanish in both polarized and unpolarized cases as shown in Appendix~\ref{appendix:perturbation}.
Therefore, second-order perturbation theory gives the exact correlation energy in that limit.
The results are computed (with Monte Carlo integration) for both the spin-polarized case and the spin-unpolarized case in a range of $r_s/d$, and all the data points are used in the global fit.

\subsection{Weakly screened limit}
When the screening is weak, ($d\rightarrow\infty$), it can be considered as a perturbation to the unscreened HEG with Coulomb interaction. In this limit, we expand it into Taylor series, 
\begin{equation}\label{eq:Taylor}
    \frac{1}{\sqrt{r^2+d^2}}=\frac{1}{d}-\frac{r^2}{2d^3}+\mathcal{O}(\frac{1}{d^5}).
\end{equation}
Substituting Eq.~(\ref{eq:Taylor}) into the properly regularized Eq.~(\ref{eq:vimgs}), i.e., Eq.~(\ref{eq:realSpace}), we find the leading order is a conditionally convergent series, and we evaluate it by summing each pair of positive and negative term of the same $m$ and obtain 
\begin{equation}
    \hat{H}^{(1)}=\frac{-N}{2d}\left[-\frac{1}{2}+\sum_{m=1}^{\infty}\frac{-1}{2m(4m^2-1)}\right]=\frac{N\mathrm{ln}(2)}{2d}.
\end{equation} 
This term is operator-free and is only a function of $d$. The leading order correction to the total energy is then readily computed by the first order perturbation $E^{(1)}=\langle\psi^{\mathrm{Coul}}|\hat{H}^{(1)}|\psi^{\mathrm{Coul}}\rangle$. Therefore, the total energy per electron 
\begin{equation}
    \epsilon = \epsilon^{\mathrm{Coul}}+\frac{\mathrm{ln}(2)}{2r_s}\mu+O(\mu^2),
\end{equation}
where the next-order quadratic term is from the second-order perturbation.

The leading-order correction for exchange energy can be evaluated in a similar way. 
Substituting Eq.~(\ref{eq:Taylor}) into Eq.~(\ref{eq:ex_convention}), we obtain $E_x^{(1)}=\langle\psi_0|\hat{H}^{(1)}|\psi_0\rangle$. Hence,
\begin{equation}
    \epsilon_x = \epsilon_x^{\mathrm{Coul}}+\frac{\mathrm{ln}(2)}{2r_s}\mu+O(\mu^3).
\end{equation}
The correlation energy in the weakly screened limit is then $\epsilon_c=\epsilon-\epsilon_x \simeq \mu^2$.

\section{Global form and fitting results}\label{sec:Global_fit}
\subsection{Unpolarized}
Our functional form for the correlation energy in the unpolarized case is:
\begin{widetext}
    \begin{equation}\label{eq:up}
        \epsilon_c(r_s,\mu,p=0)=
        \frac{\epsilon_c^{\mathrm{Coul}}(r_s,p=0)\left[1+g_1(r_s)\mu\right]+ g_{2a}(r_s)\mu^2+ g_{3a}(r_s)\mathrm{ln}(\mu+1)\mu^3}
{1+\sum_{n=1}^5 g_n(r_s)\mu^n}\,.
    \end{equation}
\end{widetext}
Here $\epsilon_c^{\mathrm{Coul}}$ is the correlation energy per electron in the absence of screening. 
The density-dependent prefactors are in the form
\begin{equation}
    g_i(r_s)=A_i/(r_s+C)+B_i/r_s,
\end{equation}
where $A_i$, $B_i$, and $C$ are fitting parameters. 
This fitting form Eq.~(\ref{eq:up}) satisfies the previously discussed asymptotic limits. In the unscreened limit, $d\rightarrow\infty$ ($\mu\rightarrow0$), it reduces to $\epsilon^{\mathrm{Coul}}_c$. When the screening is weak, the leading-order correction is quadratic in $\mu$. In the strongly screened limit, $\mu\rightarrow\infty$, it reduces to the asymptotic expression Eq.~(\ref{eq:pert_up}), obtained from perturbation analysis.
In the high-density limit ($r_s\rightarrow 0$), it still satisfies the asymptotic limits on both the weakly screened and strongly screened sides, except that we include a term that is linear in $\mu$ in the weakly screened side, a compromise made to improve the fit to the numerical data. In the dilute limit, $r_s\rightarrow\infty$, and $\epsilon_c(r_s,\mu,p=0)\rightarrow 0$. 

We apply a least-square fit for our expression to the corresponding data from a range of $r_s$ and $d/r_s$ listed in Table~\ref{tab:unpolarized_data_1} in Appendix~\ref{appendix:Ec_data}, where the total error bar is obtained from a combination of the statistical error from DMC calculations and the errors associated to the fits in the finite-time-step and finite-size extrapolation.  
Once we obtain optimal parameters to the fit through the least-squares formulation, we calculate their standard deviation using a linear approximation~\cite{Vugrin_WRR2007}. The final fitting parameters are obtained by rounding them up to a digit consistent with ten percent of their standard deviation. Fitting parameters much smaller than their standard deviation are set to zero when appropriate.

\begin{table}
\caption{\label{tab:params} Optimal fit parameters for the correlation energy per electron. Values labeled * are required by asymptotic limits. In the spin-unpolarized case, the parameter C=245 and the reduced $\chi^2=2.4$. In the spin-polarized case, the reduced $\chi^2=8.5$.}
\begin{ruledtabular}
\begin{tabular}{ccccc}
 & \multicolumn{2}{c}{$p = 0$} & \multicolumn{2}{c}{$p = 1$} \\
\cline{2-3} \cline{4-5}
 & $A_i$ & $B_i$ & $A_i$ & $B_i$\\ 
\hline
i=1& 87 & 0.494 & 1.06 & 1.9 \\ 
i=2& 106 & 0.69 & 0.13 & 2.38 \\
i=2a& $0.11$ & $-0.089$ & $-0.46$ & $-0.0378$\\
i=3& 40.6 & 0.355 & 0.5 & 2.77 \\
i=3a& $0^*$ & ${-0.5\cross B_5}^*$ & $0.021$ & $0^*$\\
i=4& $0$ & 0.0575 & 0 & 0 \\
i=5& $0^*$ & $4\cross 10^{-5}$& 0 & 0.75\\
i=6& - & - &- &$-4.1$\\
\end{tabular}
\end{ruledtabular}
\end{table}

\begin{figure*}
\includegraphics[width=1.0\textwidth]{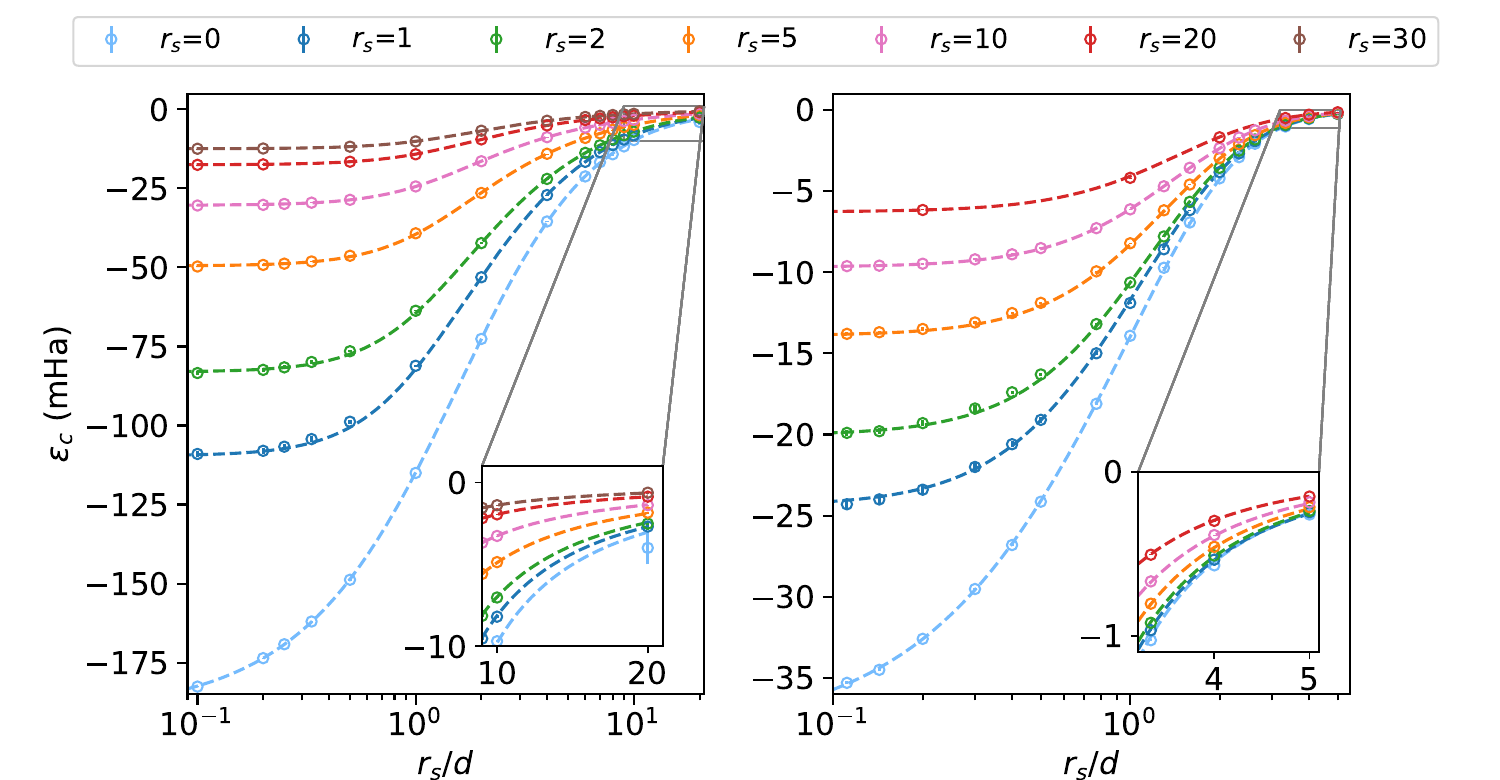}
\caption{Correlation energy $\epsilon_c$ as a function of $r_s$ and the ratio $r_s/d$ of the unpolarized (left) and polarized (right) \sys.
The open circles represent data from the FN-DMC calculations ($r_s>0$) and exact results from perturbation calculations ($r_s=0$), while the dashed curves are evaluated from the fitted analytic expressions Eq.~(\ref{eq:up}) and (\ref{eq:p}).
The insets show the approach to the non-interacting limit in linear scale. 
}
\label{fig:fitted}
\end{figure*}

\subsection{Polarized}
In the spin-polarized case, the functional form is:
\begin{widetext}
\begin{equation}\label{eq:p}
        \epsilon_c(r_s,\mu,p=1)=
\frac{\epsilon_c^{\mathrm{Coul}}(r_s,p=1)\left[1+h_1(r_s)\mu\right]
+ h_{2a}(r_s)\mu^2 + h_{3a}(r_s)\mu^3}
{1+\sum_{n=1}^6 h_n(r_s)\mu^n}.
\end{equation}     
\end{widetext}
The density-dependent prefactors are given by: 
\begin{align}
\begin{split}
    &h_i(r_s)=\begin{cases}
    A_i+B_i/r_s,&i=1...5\\\    
    A_{i}\epsilon_c^{\mathrm{Coul}}(r_s,p=1)+B_{i}/r_s, &i=2a,3a\\
    B_{i}\epsilon_c^{\mathrm{Coul}}(r_s,p=1)/r_s,              & i=6\\
    \end{cases}\\
\end{split}
\end{align}
where $A_{i}$ and $B_{i}$ are fitting parameters. The fitting form Eq.~(\ref{eq:p}) satisfies the same asymptotic limits that apply to Eq.~(\ref{eq:up}) except that in the strongly screened limit $\mu\rightarrow\infty$, where it reduces to the asymptotic expressions, Eq.~(\ref{eq:pert_p1}), and Eq.~(\ref{eq:pert_p2}), obtained from perturbation analysis.

Similar to the spin-unpolarized case, we apply a least-square fit for our expression to the data from a range of $r_s$ and $d/r_s$ listed in Table~\ref{tab:polarized_data_1}. 

The optimal fitting parameters are listed in Table~\ref{tab:params}. The fitted results are plotted with the data points in Fig.~\ref{fig:fitted}.

\subsection{Intermediate polarization}

\begin{figure}
\includegraphics[width=0.5\textwidth]{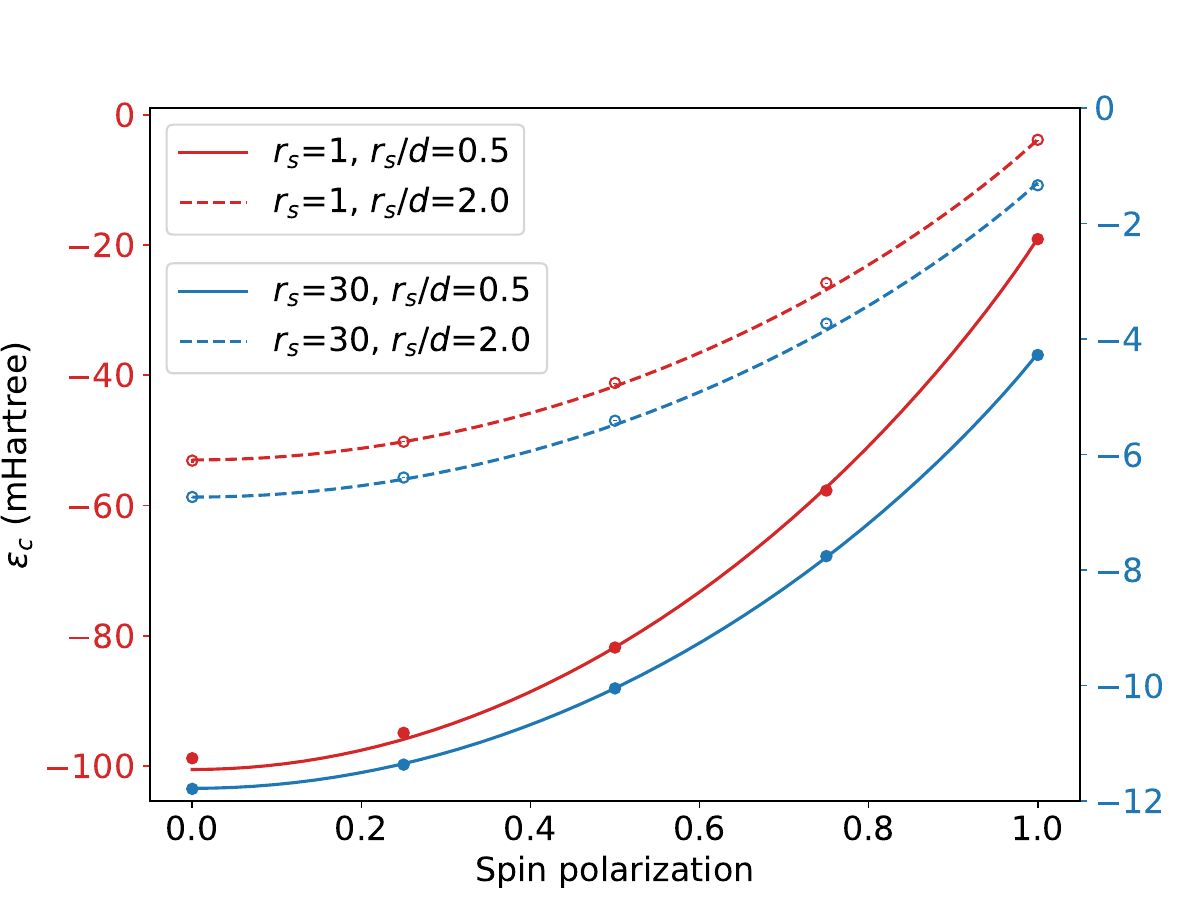}
\caption{Quality of interpolated correlation energy $\epsilon_c$ at intermediate polarization.
The curves are interpolated values from Eq.~(\ref{eq:interpolate}).
The symbols are FN-DMC results obtained at intermediate polarizations (not included in the fitting of $\epsilon_c$).
Two densities are shown with different y-axis: red (left) for $r_s=1$ and blue (right) for $r_s=30$.
At each density, two screening strengths $\mu=0.5, 2.0$ are presented as the solid and dashed curves, respectively.
The relative error in the interpolation is within $5$ percent, in all cases shown.
}
\label{fig:lsda}
\end{figure}

To obtain the partially spin-polarized results, we apply an exchange-like expression~\cite{Koskinen_PRL1997,Gori-Giorgi_IJQC2003} to interpolate the spin-polarized and spin-unpolarized results:
\begin{equation}\label{eq:interpolate}
\begin{split}
\epsilon_c(r_s,\mu,p)=\epsilon_c(r_s,\mu,0)+&\frac{P_+^{3}+P_-^{3}-2}{2^{3/2}-2}\\
 &\cross[\epsilon_c(r_s,\mu,1)-\epsilon_c(r_s,\mu,0)],
\end{split}
\end{equation}
where $P_+$ and $P_-$ are the same as that appear in Eq.~(\ref{eq:bc_HF}).
We performed DMC calculations of partially spin-polarized systems at $r_s=1$ and $30$ with the ratio $r_s/d=0.5$ and $2$ to verify the quality of this interpolation. Results are plotted together with the interpolation in Fig.~\ref{fig:lsda}. The relative error between the DMC calculations and the simple interpolation is within $5$ percent, suggesting that this approximation is reasonable in the presence of screening.

\section{Other properties}\label{sec:Other_properties}

\subsection{Bulk modulus}
\begin{figure}
\includegraphics[width=0.5\textwidth]{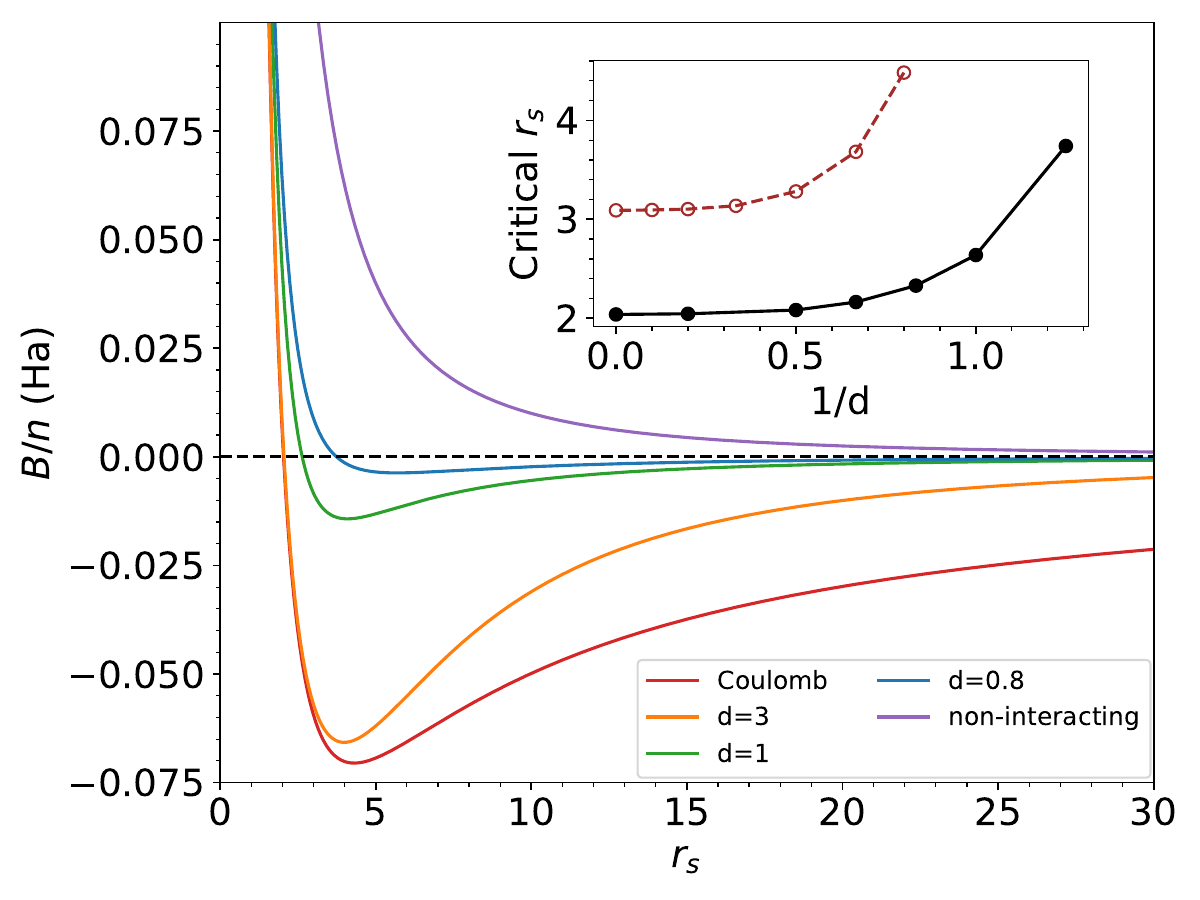}
\caption{Bulk modulus of the \sys~as a function of $r_s$ at fixed gate distance.
The main plot shows how the bulk modulus over density $B/n$ vs. $r_s$ changes with screening strength in the spin-unpolarized case. The critical $r_s$ is the value where $B/n$ changes sign. The inset shows how the critical $r_s$ varies, as a function of screening strength, tuned by the gate distance. The solid black curve and dashed brown curve show the unpolarized and polarized critical $r_s$ as a function of the gate distance, respectively.
}
\label{fig:B}
\end{figure}

Using the parameterized correlation energy, the bulk modulus $B(r_s)=-V(\partial p/\partial V)_N$ can be readily computed, where $V$ is the volume and $p$ is the pressure.
The inverse of $B(r_s)$ can be measured by probing the electric fields penetrated through a two-dimensional electron system~\cite{Eisenstein_PRL1992}. $B(r_s)$ changes sign at a critical $r_s$ as the electron density decreases. Analytically, it can be evaluated from the derivatives of the total energy per electron~\cite{Giuliani_Book2005} $\epsilon(r_s)$: $B(r_s)/n=r_s(r_s \epsilon''(r_s)-\epsilon'(r_s))/4 $, where the prime and double prime in $\epsilon$ denote the first and second derivative with respect to $r_s$, respectively. We find that the critical $r_s$ increases dramatically with the screening brought by the gates and that the $B(r_s)$ approaches the non-interacting results, as shown in Fig.~\ref{fig:B}.

\subsection{Pair-correlation function}

\begin{figure*}
\includegraphics[width=1\textwidth]{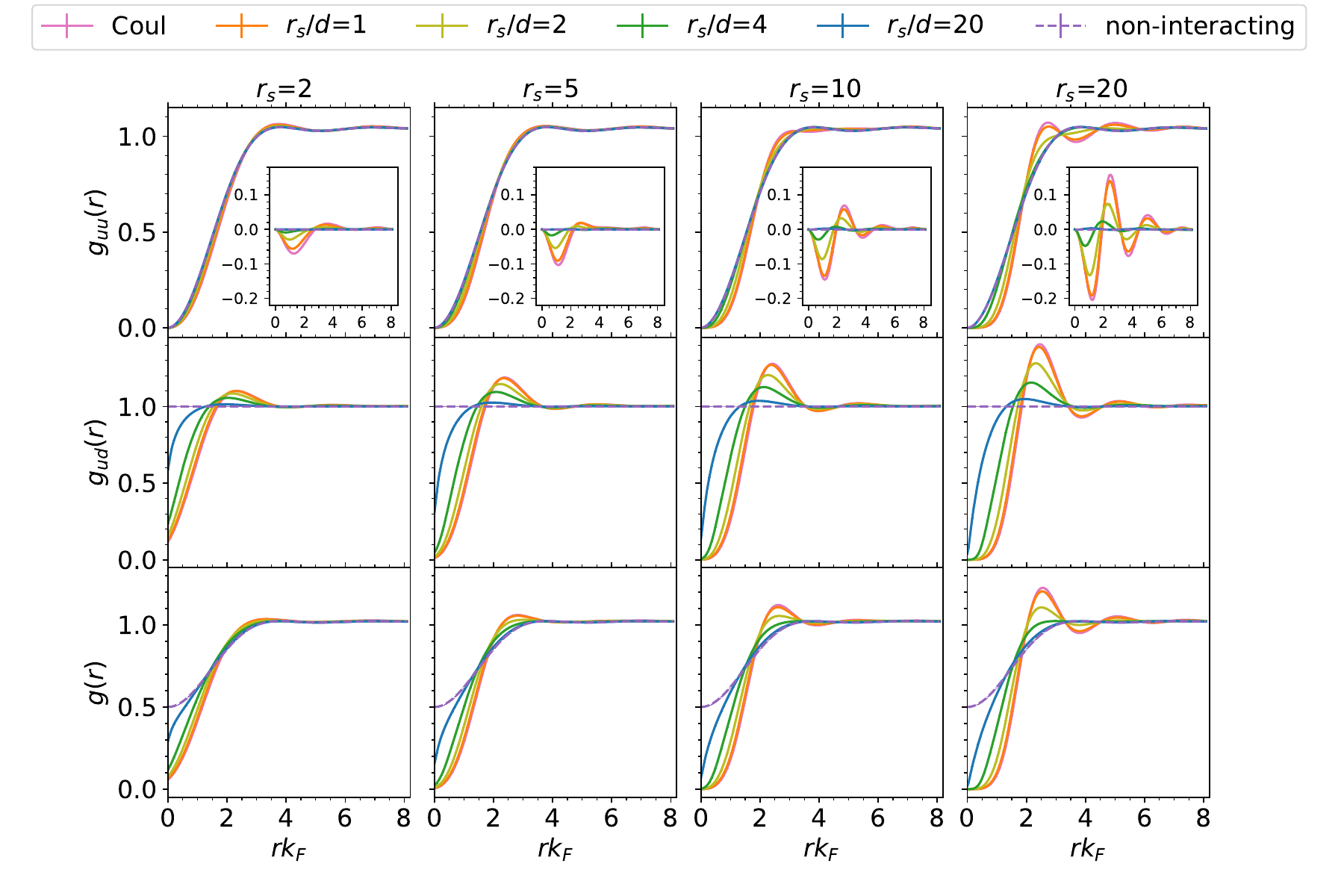}
\caption{Spin-resolved pair correlation functions of the unpolarized \sys~ 
for at several $r_s$ and $\mu$.
Each column corresponds to one density, while the color encodes the screening strength.
The top, middle, and bottom rows show the same-spin, opposite-spin, and total pair correlation functions, respectively.
The insets show the difference relative to the non-interacting limit (dashed purple curves).
}
\label{fig:gr0p}
\end{figure*}

In addition to the total energy, we also computed the spin-resolved pair correlation function, defined as 
\begin{equation}
g_{\sigma\sigma'}(\mathbf{r})=\int_{A}d \mathbf{r}'\frac{\langle \hat{n}(\mathbf{r}')_{\sigma}\hat{n}_{\sigma'}(\mathbf{r'+r})\rangle}{\langle \hat{n}(\mathbf{r}')_{\sigma}\rangle \langle \hat{n}_{\sigma'}(\mathbf{r'+r})\rangle},
\end{equation}
where $\hat{n}_{\sigma}$ is the electron density operator with spin $\sigma$ and $A$ denotes the area of our simulation cell. Since our system is isotropic, we averaged $g_{\sigma\sigma'}(\mathbf{r})$ on different directions and obtained $g_{\sigma\sigma'}(r)$.
The spin-resolved pair correlation functions of different $r_s$ and gate distances in the spin-unpolarized case are shown in Fig.~\ref{fig:gr0p}. The corresponding results for the bare Coulomb interaction and the non-interacting cases are also plotted for comparison. The correlation energy vanishes in the absence of interaction, but parallel-spin electrons are still correlated~\cite{Gori-Giorgi_PRB2004,Giuliani_Book2005}:
\begin{equation}
    g_{\sigma\sigma}^x=1-\left[\frac{2J_1(k_F^{\sigma}r)}{k_F^{\sigma}r}\right]^2
\end{equation}
This is a feature for fermions that are described by the Fermi-Dirac statistics, a consequence of the Friedel oscillation~\cite{Ashcroft_Book1976}. The region where electrons of the same spin are less likely to appear is called the exchange hole. In contrast, the non-interacting anti-parallel-spin correlation function $g^x_{ud}(r)=1$, which is independent of $r$, indicating the non-interacting anti-parallel-spin electrons are uncorrelated. $g^x_{\sigma\sigma}$ is the same for different electron densities when scaled by $k_F$.

\begin{figure*}
\includegraphics[width=1\textwidth]{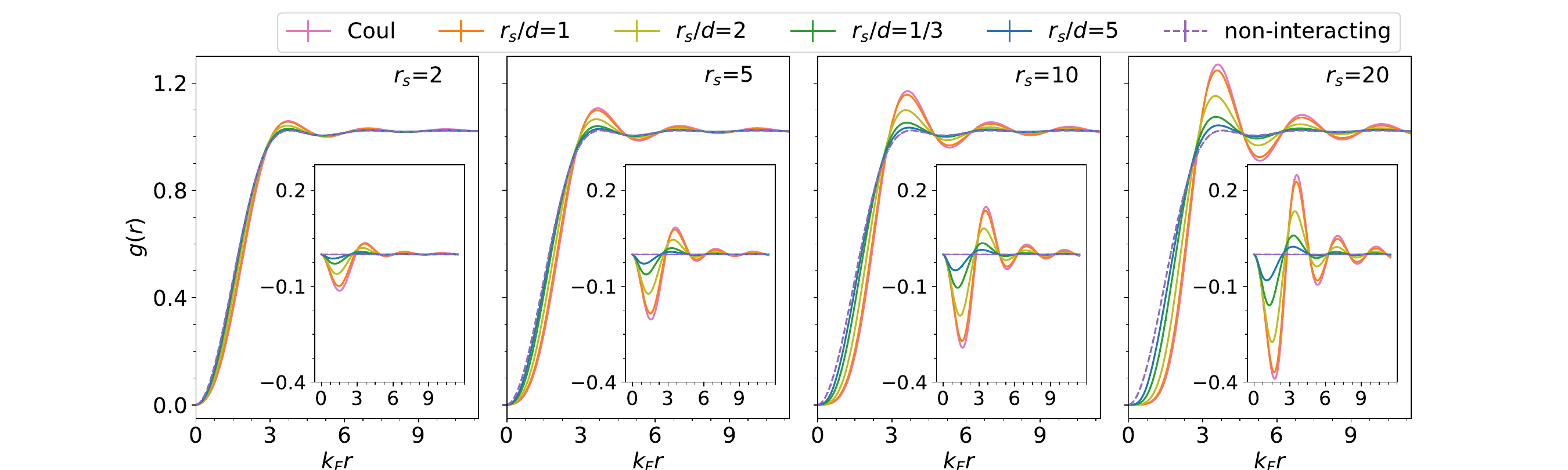}
\caption{Pair correlation functions of the polarized \sys, similar to Fig.~\ref{fig:gr0p}.}
\label{fig:gr1p}
\end{figure*}

With interaction, the correlation between both parallel- and antiparallel-spin electrons increases with $r_s$, as is reflected by the expansion of the depletion region (exchange-correlation hole) on the left of each panel. However, gate screening dials down the enhanced correlation in a more dilute HEG. As the gate distance decreases, the pair correlation function approaches the non-interacting limit $g^x_{\sigma\sigma}$. The influence of the gate screening to $g_{\sigma\sigma'}(r)$ becomes increasingly pronounced as $r_s$ increases, indicating the growing importance of the electron-electron interactions. For antiparallel-spin electrons, the depletion region is called the correlation hole due to the absence of the exchange effect. It's radius is significantly smaller than the exchange-correlation hole for parallel spin electrons, which makes $g_{ud}(r)$ more sensitive to the gate-screening than $g_{uu}(r)$. The on-top pair density $g_{ud}(0)$ is finite for anti-parallel electron pairs because they are not affected by the Pauli exclusion principle. We can easily see from the second row, for instance in $r_s=2$ and $5$, the on-top pair density increases with the gate screening. The correlation functions in the spin-polarized case are similar, as shown in Fig.~\ref{fig:gr1p}. 

\subsection{Static structure factor}
The static structure factor $S(k)$:
\begin{equation}
    S(\mathbf{q})=1+n\int[g(\mathbf{r})-1]e^{-i\mathbf{q}\cdot\mathbf{r}}d\mathbf{r}
\end{equation}
is the Fourier transform of the pair correlation function and can be measured by X-ray scattering. Without screening, it is well known that when the density is low, the dominant electron-electron interaction leads to the formation of a Wigner crystal, where $S(k)$ contains sharp peaks (Bragg peaks). In contrast, when the density is high, the kinetic energy dominates and the peaks vanish. At intermediate densities, the $S(k)$ in a 2D HEG exhibits a broad peak, indicating the presence of short-range charge-charge correlation. 

\begin{figure}
\includegraphics[width=0.5\textwidth]{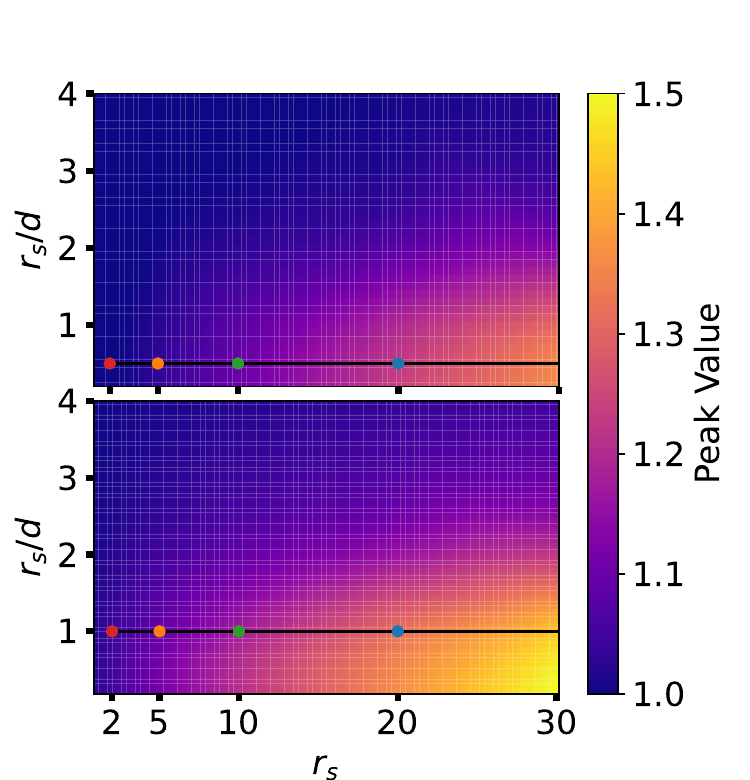}
\caption{Heat plots of the magnitude of the peak in $S(k)$ of the 2D HEG under dual-gate screening under a range of $r_s$ and $r_s/d$. Necessary interpolations are carried out. The top panel is for the spin-unpolarized case and the bottom panel is for the spin-polarized case. The horizontal solid black lines correspond to fixed $r_s/d$. The full $S(k)$ of the color markers are shown in Fig.~\ref{fig:lineCuts}.
}
\label{fig:heatMap}
\end{figure}

\begin{figure}
\includegraphics[width=0.5\textwidth]{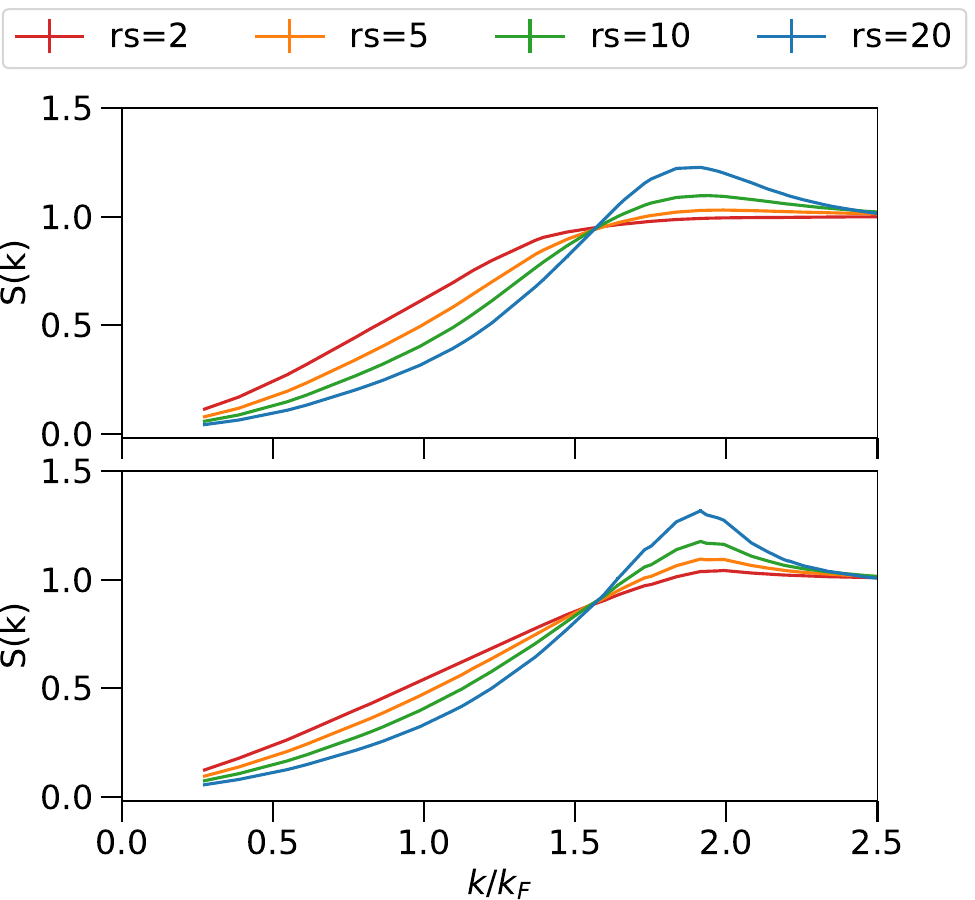}
\caption{The full $S(k)$ of the color markers in Fig.~\ref{fig:heatMap}.  The top row is for $r_s/d=0.5$ in the spin-unpolarized case and the bottom row is for $r_s/d=1$ in the spin-polarized case. 
}\label{fig:lineCuts}
\end{figure}

The presence of metallic gates suppresses the broad peak in the structure factor. At a fixed $r_s$, the peak decreases monotonically with the screening strength $\mu$. On the other hand, at fixed $\mu$, the peak is enhanced by the interaction strength, increasing monotonically with $r_s$. The same qualitative behavior is observed in both spin-polarized and spin-unpolarized systems, as shown in Fig.~\ref{fig:heatMap}. The horizontal solid black lines represent varying the electron density under a fixed $\mu$. We plot $S(k)$ for the four $r_s$ values marked by circles of different colors in Fig.~\ref{fig:lineCuts} as an example.

At a fixed gate distance, $\mu$ increases linearly with $r_s$, leading to a competition between gate screening and interaction effects. We observed that the broad peak in $S(k)$ tends to initially increase but eventually decrease with $r_s$. This indicates that the electron density plays a major role in correlation when $r_s$ is small while gate screening takes over when $r_s$ is large. This behavior can be attributed to the fact that, at lower densities, electrons are more widely separated, and the dual gate more effectively screens interactions at longer distances compared to shorter distances (see Appendix~\ref{appendix:gateScreening}). In experiments, since the gate distance $d$ is typically fixed, this phenomenon could be observed.

Some more line cuts of the heat plots for a range of $r_s$ under different screening are plotted in Fig.~\ref{fig:sk0p} and ~\ref{fig:sk1p} in Appendix~\ref{appendix:Sk}.

\section{Conclusions}\label{sec:Conclusions}
We have created an accurate exchange-correlation functional for the  2D HEG in the presence of dual metallic gates. QMC calculations are performed systematically as a function of the electron density, spin-polarization, and the gate distance. 
A global parametrization is obtained for the correlation energy, which accurately reproduces the QMC calculations on a large range of densities and spin polarizations consistent with the liquid state in the unscreened system. 
Furthermore, we compute the bulk modulus, pair correlation function, and structure factor from QMC and study their dependence on gate separation.

Advances in the study and understanding 2D materials have been explosive, leading to many new and different moir\'e systems being realized and investigated experimentally. 
Most experimental setups involve metallic gates, which can have highly non-trivial effects on the properties of the system. DFT calculations are widely used as an important tool to both gain insight into the microscopic properties of the materials and to help guide experimental setups and designs.
To date, these studies have been hindered by the lack of accurate exchange-correlation functionals in 2D for screened interactions. 
The exchange-correlation functional presented in this work goes a long way toward eliminating this limitation, enabling DFT studies of realistic models of 2D materials with an accurate treatment of correlation effects in the presence of gate screening. 

Our results provide valuable insights into the effect of screening and its interplay with correlation effects in the 2D HEG. 
The correlation functions and other properties, obtained from accurate many-body computations, will enhance our theoretical understanding and serve as a benchmark for future calculations. They can also potentially be probed by direct experimental measurements.

\section{Acknowledgements}
We thank Pengcheng Hou for useful discussions on perturbation analysis
and the
Flatiron Institute for hospitality and computational resources. The Flatiron Institute is a division of the Simons Foundation.

\clearpage
\bibliography{ref}

\clearpage
\onecolumngrid
\renewcommand{\thefigure}{A\arabic{figure}}
\renewcommand{\thetable}{A\arabic{table}}
\setcounter{figure}{0}
\setcounter{table}{0}
\appendix
\section{The Coulomb potential under symmetric dual-gate screening}\label{appendix:gateScreening}
\begin{figure}[h]
\centering
\includegraphics[scale=0.2]{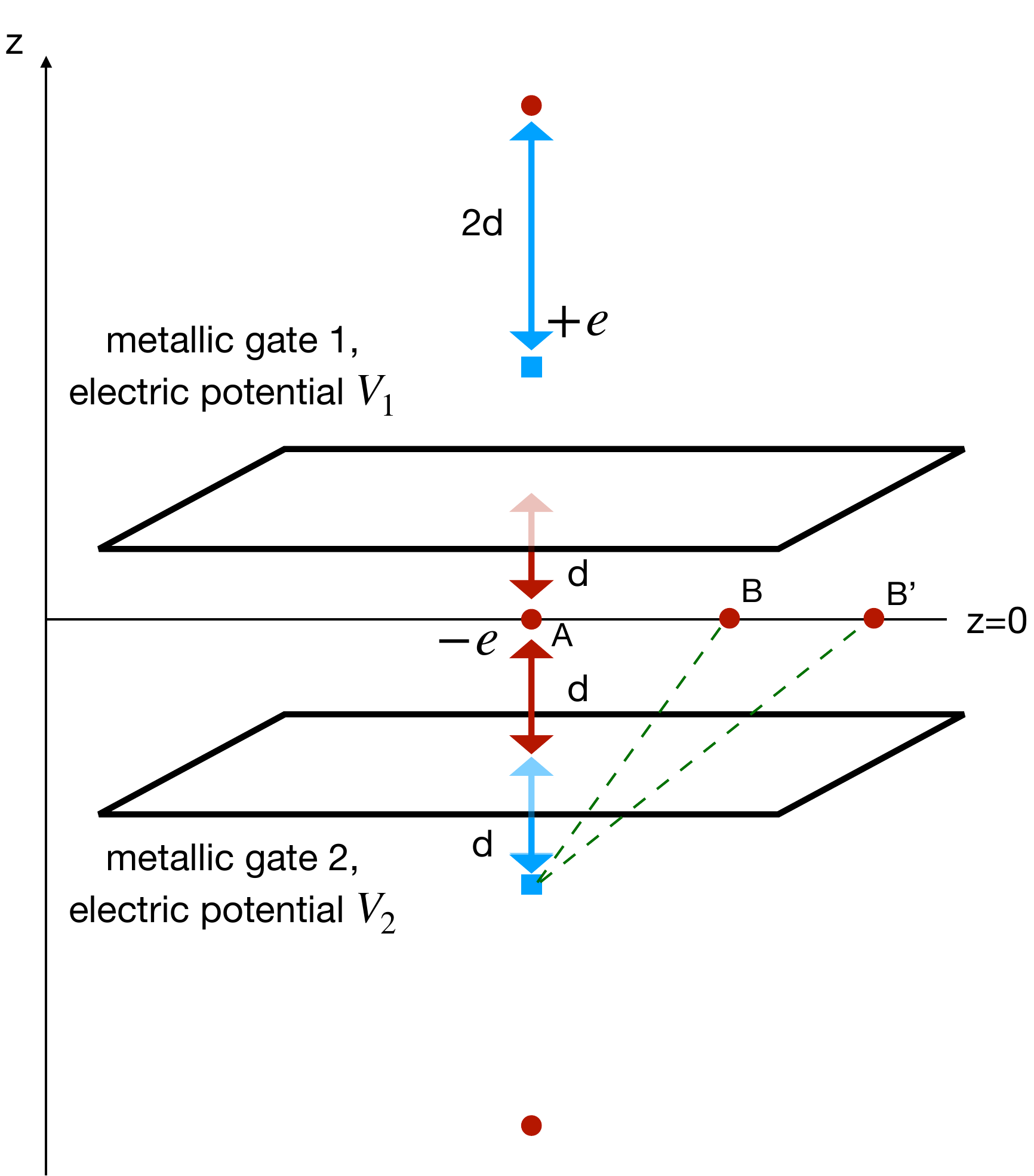}
\caption{Schematic representation of gate screening effects. Each metallic gate maintains a uniform electric potential. An electron labeled A is positioned equidistant $d$ from both gates in the middle plane between them. The other red dots in the middle plane are other electrons in the 2D HEG in the $z=0$ plane. The dots outside of the middle region of the two gates are image charges. Blue squares represent positive charges ($+e$) and red circles represent negative charges ($-e$). The dashed green lines denote the distance between electron $B$, $B'$ and the first image charge in the $z<0$ region, respectively.}
\label{fig:gateScreening}
\end{figure}

The metallic gates impose a boundary condition on the system, requiring that the electric fields vanish within the metallic materials. This boundary condition uniquely determines the electric fields between the two gates. By treating each gate as an infinite plane, we can solve the electric fields in the region between them. This is achieved by replacing the two metallic gates with an infinite array of image charges, as illustrated in Fig.~\ref{fig:gateScreening}.
The electric potential between the two gates is now equal to the sum of the potentials generated by the charge in the middle and the infinite array of image charges. The coordinates of the positive image charges are given by $z=d_m^+=(4m-2)d$, where $m$ is an integer ranging from $-\infty$ to $\infty$.  In contrast, the coordinates of the negative image charges are $z=d_m^-=4md$. Consequently, the electric potential in the middle plane between the two gates can be expressed as follows:
\begin{equation}\label{eq:realSpace}
    v(r)=\sum_{m=-\infty}^{+\infty}\left[(r^2+{d_m^-}^2)^{-1/2}-(r^2+{d_m^+}^2)^{-1/2}\right].
\end{equation}
The Fourier transform of each term in Eq.~(\ref{eq:realSpace}) can be written as:
\begin{equation}
    \int d^2\mathbf{r}\frac{e^{-ikr}}{\sqrt{r^2+c^2}}=\frac{2\pi e^{-\abs{c}k}}{k}.
\end{equation}
Therefore, in the momentum space, the electric potential:
\begin{equation}\label{eq:geo_series}
    v(k)=\frac{2\pi }{k}\sum_{m=-\infty}^{+\infty}\left[e^{-\abs{d_m^-}k}-e^{-\abs{d_m^+}k}\right].
\end{equation}
When $k\neq 0$, it is a geometric series and has a compact form:
\begin{equation} \label{eq:tanh}
    v(k)=\frac{2\pi \mathrm{tanh}(dk)}{k}.
\end{equation}
When $k=0$, both Eq.~(\ref{eq:geo_series}) and Eq.~(\ref{eq:tanh}) vanish. 

The gates significantly screen Coulomb interactions between electrons that are separated more than the gate distance. Taking the $m=0$ term in Eq.~(\ref{eq:realSpace}) for example, only the first image charge in the $z<0$ space appears in this term. When the distance between electron $A$ and electron $B$ in the $z=0$ plane increases ($B\rightarrow B'$), it approaches the distance between electron $B$ and the image charge (denoted by the dashed green lines), as is shown in Fig.~(\ref{fig:gateScreening}). This leads to a larger cancelation for interactions in the plane. The same argument applies to each term $m$ in Eq.~(\ref{eq:realSpace}).

\section{Fit the correlation energy of \sys~with bare Coulomb interaction}\label{appendix:EcBC}
Drummond et. al~\cite{Drummond_PRL2009} reported that in the spin-unpolarized case the crystal phase shows up at $r_s=31 (1)$ while in the spin-polarized case this happens between $r_s=25$ and $r_s=30$. We generated data in the fluid phase region accordingly with Coulomb interaction (no gate screening) as shown in Table.~\ref{tab:bc_data}.
\begin{table}
\centering
\caption{Correlation energy per electron in the fluid phase of \sys in both the spin-unpolarized case and spin-polarized case obtained from QMC.}\label{tab:bc_data}
\setlength{\tabcolsep}{1pt}
\begin{ruledtabular}  
\begin{tabular}{ccc}
      &$p=0$ & $p=1$ \\
      \hline
$r_s=1$ & -0.1096(2)	  &-0.0244(1)\\
$r_s=2$ & -0.0832(2)	  &-0.02006(8)\\
$r_s=3$ & -0.0674(2)	  &-0.01733(6)\\
$r_s=4$ & -0.0570(2)	  &-0.01539(5)\\
$r_s=5$ & -0.04958(7)	  &-0.01392(4)\\
$r_s=6$ & -0.0438(1)	  &-0.01275(4)\\
$r_s=8$ & -0.0358(1)	  &-0.01098(3)\\
$r_s=10$ & -0.03036(3)	  &-0.00970(2)\\
$r_s=15$ & -0.02214(5)	  &-0.00759(2)\\
$r_s=20$ & -0.01754(2)	  &-0.00629(1)\\
$r_s=25$ & -0.01455(3)	  &-0.00540(1)\\
$r_s=30$ & -0.01247(1)	 & $\cdots$
\end{tabular}
\end{ruledtabular}
\end{table}
We fit them in the spin-polarized case and spin-unpolarized case, respectively using Attaccalite's fitting expression~\cite{Attaccalite_PRL2002} in the limit of zero polarization:
\begin{equation} \label{eq:at-ec-p0}
\begin{split}
    \epsilon_c^{\mathrm{Coul}}(r_s,p=0)=\alpha_{i=0}(r_s)
\end{split}
\end{equation}
and full polarization:
\begin{equation}
    \begin{split}
    \epsilon_c^{\mathrm{Coul}}(r_s,p=1)=&(e^{-\beta r_s}-1)\bigg[\epsilon_x(r_s,1)-\frac{179}{128}\epsilon_x(r_s,0)\bigg]+\sum_{i=0}^3 \alpha_i(r_s),
    \end{split}
\end{equation}
where $\alpha_i(r_s)$ is defined as:
\begin{equation}  \label{eq:at-ec-alpha}
\begin{split}
    \alpha_i(r_s)=&A_i+(B_ir_s+C_ir_s^2+D_ir_s^3)\times\mathrm{ln}\left(1+\frac{1}{E_ir_s+F_ir_s^{3/2}+G_ir_s^2+H_ir_s^3}\right).
\end{split}
\end{equation}
The optimal parameters we obtained from the fit is summarized in  Table~\ref{tab:bc_param}.
\begin{table}[b]
\centering
\caption{Optimal fitting parameters for the correlation energy in the unscreened case. Values labeled * are required by asymptotic limits. The parameter $\beta=1.2409$ and $D_i=-A_i*H_i$ are not listed. The reduced $\chi^2=0.3$ in the spin-unpolarized case and $\chi^2=0.1$ in the spin-polarized case.}\label{tab:bc_param}
\begin{ruledtabular}  
\begin{tabular}{cccc}
      &$i=0$ & $i=1$ & $i=2$ \\
      \hline
$A_i$ & $-0.1912^*$ & $0.117331^*$ & $0.0234188^*$ \\
$B_i$ & $0.0863136^*$ & $-0.03051$& $-0.037093^*$ \\
$C_i$ & $0.0387$ & ${-7.66765\times 10^{-3}}^*$ &$0.0163618^*$ \\
$E_i$ & $0.9308$& $0.383$& $1.3825$\\
$F_i$ & $-0.093$& $0^*$& $0^*$ \\
$G_i$ & $0.2948$& $0.08363$& $0^*$\\
$H_i$ & $0.0367$& $0.00927$& $2.236$\\
\end{tabular}
\end{ruledtabular}
\end{table}
From perturbation theory, we obtained $\epsilon_c^{\mathrm{Coul}}(0,p=0)=-0.1912(8)$ and $\epsilon_c^{\mathrm{Coul}}(0,p=1)=-0.0387(3)$, which slightly differ from previously reported values~\cite{Seidl_PRB2004}. We use these new values (the mean) as constraints in the fit. The adjustment is reflected in the changes to the parameters $\beta$ and $A_0$ compared to those fixed by exact conditions in Attaccalite's fit~\cite{Attaccalite_PRL2002}.

Next, we study the influence of not having crystal phase data in the fit. We found QMC data of the total energy in the crystal phase from Ref.~\cite{Drummond_PRL2009}. We subtract the restricted HF energy (fluid phase) from the total energy to obtain the correlation energy~\cite{Martin_Book2016}. We compare them to our fitted results and compute the relative error. The errors are found to be small, as summarized in Table~\ref{tab:crystal_data_1} and Table~\ref{tab:crystal_data_2}:
\begin{table*}[b]
\caption{Relative error of the fitted results and the correlation energy of the crystal phase in the spin-unpolarized case computed by Drummond et. al~\cite{Drummond_PRL2009}. The data with * is taken directly from Ref.~\cite{Drummond_PRL2009}. $\epsilon$ denotes the total energy, $\epsilon_{\mathrm{HF}}$ denotes the HF energy, $\epsilon_c$ denotes the correlation energy, and $\epsilon_c^{\mathrm{fit}}$ correlation energy from the fit. They are all averaged on each electron. Results from the fit are rounded to the same digit of Drummond's data.
\label{tab:crystal_data_1}}
\begin{ruledtabular}
\begin{tabular}{cccccc}
$r_s$ &$\epsilon$ &$\epsilon_{\mathrm{HF}}$ &$\epsilon_c$ &$\epsilon_c^{\mathrm{fit}}$ &$(\epsilon_c^{\mathrm{fit}}-\epsilon_c)/|\epsilon_c|\times100\%$\\
\hline
$35$ &$-0.027669(1)^*$ &-0.016740719 &-0.010928(1) &-0.010912 &$0.1\%$\\
$40$ &$-0.024432(1)^*$ &-0.014692772 &-0.009739(1) &-0.009711 &$0.3\%$\\
$45$ &$-0.021875(2)^*$ &-0.0130911 &-0.008784(2) &-0.008754 &$0.3\%$\\
$50$ &$-0.019814(2)^*$ &-0.0118042 &-0.008010(2) &-0.007971 &$0.5\%$\\
\end{tabular}
\end{ruledtabular}
\end{table*}

\begin{table*}[b]
\caption{Relative error of the fitted results and the correlation energy of the crystal phase in the spin-polarized case computed by Drummond et. al~\cite{Drummond_PRL2009}. The data with * is directly taken from Ref.~\cite{Drummond_PRL2009}.$\epsilon$ denotes the total energy, $\epsilon_{\mathrm{HF}}$ denotes the HF energy, $\epsilon_c$ denotes the correlation energy, and $\epsilon_c^{\mathrm{fit}}$ correlation energy from the fit. They are all averaged on each electron. Results from the fit are rounded to the same digit of Drummond's data.
\label{tab:crystal_data_2}}
\begin{ruledtabular}
\begin{tabular}{cccccc}
$r_s$ &$\epsilon$ &$\epsilon_{\mathrm{HF}}$ &$\epsilon_c$ &$\epsilon_c^{\mathrm{fit}}$ &$(\epsilon_c^{\mathrm{fit}}-\epsilon_c)/|\epsilon_c|\times100\%$\\
\hline
$30$ &$-0.031917(2)^*$ &-0.027183101 &-0.004734(2) &-0.004733 &$0.01\%$\\
$35$ &$-0.027669(1)^*$ &-0.023435855 &-0.004233(1) &-0.004224 &$0.2\%$\\
$40$ &$-0.024432(1)^*$ &-0.020595659 &-0.003836(1) &-0.003818 &$0.5\%$\\

\end{tabular}
\end{ruledtabular}
\end{table*}

\COMMENTED{
As we finish this work, we found that Azadi et al.~\cite{Azadi_arxiv2024} reported the QMC data of \sys in the fluid phase. We plotted a comparison between our data and that obtained by Azadi et al. in Fig.\ref{fig:compare}.
\begin{figure}
\includegraphics[scale=0.5]{compare.pdf}
\centering
\caption{Compare the correlation energy in the fluid phase in this work with that reported by Azadi et al.~\cite{Azadi_arxiv2024} for a range of $r_s$ in both the spin-polarized and spin-unpolarized cases. In the y-label, $\epsilon_c^{A}$ denotes the correlation energy per electron reported by Azadi et al. and $\epsilon_c$ denotes that from the present work.}
\label{fig:compare}
\end{figure}
}

\section{Correlation energy of \sys~under dual-gate screening}\label{appendix:Ec_data}
The data we used to fit the spin-unpolarized case is presented in Table~\ref{tab:unpolarized_data_1}, while the data used to fit the spin-polarized case is shown in Table~\ref{tab:polarized_data_1}. Data for $r_s=0$ is obtained from perturbation theory, and data for other finite $r_s$ is from QMC calculations. Additionally, data for partially spin-polarized systems is provided in Table~\ref{tab:partially_spin_polarized}.
\begin{table*}[b]
\caption{Correlation energy per electron ($\epsilon_c$) for the spin unpolarized \sys with double-gate screening. $\epsilon_c$ at $r_s=0$ are calculated from perturbation and all other data are quantum Monte Carlo (QMC) results. The unit is Hartree. The data are all under thermodynamic limit. \label{tab:unpolarized_data_1}}
\begin{ruledtabular}
\begin{tabular}{cccccccc}
&$r_s=0$ &$r_s=1$ &$r_s=2$ &$r_s=5$ &$r_s=10$ &$r_s=20$ &$r_s=30$\\
\hline
$d/r_s=0.05$ &-0.004(1)	  &-0.0027(2)	  &-0.00252(8)	  &-0.00185(6)	  &-0.001393(5)	  &-0.000874(4)	  &-0.000630(4)\\
$d/r_s=0.1$ &-0.00970(7)	  &-0.00820(7)	  &-0.00703(4)	  &-0.00487(3)	  &-0.003272(7)	  &-0.001954(3)	  &-0.001394(2)\\
$d/r_s=0.111$ &-0.0117(1)	  &-0.00954(6)	  &-0.00815(4)	  &-0.00559(3)	  &-0.003689(7)	  &-0.002190(3)	  &-0.001556(2)\\
$d/r_s=0.125$ &-0.0142(5)	  &-0.01128(6)	  &-0.00950(4)	  &-0.00643(3)	  &-0.004217(7)	  &-0.002485(3)	  &-0.001766(2)\\
$d/r_s=0.143$ &-0.01670(7)	  &-0.01351(7)	  &-0.01134(5)	  &-0.00758(3)	  &-0.004895(9)	  &-0.002870(4)	  &-0.002033(2)\\
$d/r_s=0.167$ &-0.0212(2)	  &-0.01659(8)	  &-0.01376(5)	  &-0.00904(3)	  &-0.00576(2)	  &-0.003373(4)	  &-0.002387(3)\\
$d/r_s=0.25$ &-0.03545(7)	  &-0.0271(1)	  &-0.02199(3)	  &-0.01404(1)	  &-0.00879(1)	  &-0.005071(6)	  &-0.003581(3)\\
$d/r_s=0.5$ &-0.07261(9)	  &-0.0531(3)	  &-0.0423(2)	  &-0.02643(3)	  &-0.01640(2)	  &-0.00949(1)	  &-0.006736(8)\\
$d/r_s=1.0$ &-0.1150(1)	  &-0.0811(4)	  &-0.0637(2)	  &-0.0392(1)	  &-0.02437(5)	  &-0.01421(2)	  &-0.01013(2)\\
$d/r_s=2.0$ &-0.1488(3)	  &-0.0988(7)	  &-0.0765(4)	  &-0.0463(2)	  &-0.0286(1)	  &-0.01655(5)	  &-0.01179(3)\\
$d/r_s=3.0$ &-0.1620(1)	  &-0.1043(9)	  &-0.0799(6)	  &-0.0481(2)	  &-0.0295(1)	  &$\cdots$       &$\cdots$\\
$d/r_s=4.0$ &-0.1692(2)	  &-0.1067(9)	  &-0.0816(6)	  &-0.0488(3)	  &-0.0299(1)	  &$\cdots$       &$\cdots$\\
$d/r_s=5.0$ &-0.1736(2)	  &-0.108(1)	  &-0.0824(7)	  &-0.0492(3)	  &-0.0302(1)		&-0.01740(7)	  &-0.01237(5)\\
$d/r_s=10.0$ &-0.1826(2)	  &-0.109(1)	  &-0.0834(7)	  &-0.0497(3)	  &-0.0304(1)	&-0.01754(8)	  &-0.01245(5)\\
\end{tabular}
\end{ruledtabular}
\end{table*}

\begin{table*}[b]
\caption{Correlation energy per electron ($\epsilon_c$) for the spin polarized \sys with double-gate screening. $\epsilon_c$ at $r_s=0$ are calculated from perturbation and all other data are QMC results. The unit is Hartree. The data are all under thermodynamic limit. \label{tab:polarized_data_1}}
\begin{ruledtabular}
\begin{tabular}{ccccccc}
&$r_s=0$ &$r_s=1$ &$r_s=2$ &$r_s=5$ &$r_s=10$ &$r_s=20$\\
\hline
$d/r_s=0.2$ &-0.00026(4)	  &-0.000246(1)	  &-0.000238(4)	  &-0.000215(3)	  &-0.000185(2) &-0.00015(2)\\
$d/r_s=0.25$ &-0.00057(6)	  &-0.000535(2)	  &-0.00051(3)	  &-0.000457(3)	  &-0.000385(2) &-0.0002985(9)\\
$d/r_s=0.3$ &-0.001027(6)	  &-0.000966(4)	  &-0.000921(3)	  &-0.000804(2)	  &-0.000668(2) &-0.000505(2)\\
$d/r_s=0.38$ &-0.00209(1)	  &-0.001937(7)	  &-0.001827(6)	  &-0.00156(4)	  &-0.001265(3) &$\cdots$ \\
$d/r_s=0.43$ &-0.00292(1)	  &-0.002682(9)	  &-0.002517(7)	  &-0.002122(6)	  &-0.001706(5) &$\cdots$\\
$d/r_s=0.5$ &-0.00422(1)	  &-0.00384(1)	  &-0.00357(1)	  &-0.002976(9)	  &-0.002355(6) &-0.001694(5)\\
$d/r_s=0.63$ &-0.00694(3)	  &-0.00618(2)	  &-0.00567(2)	  &-0.0046(1)	  &-0.00357(1)  &$\cdots$\\
$d/r_s=0.77$ &-0.00973(3)	  &-0.00859(4)	  &-0.0078(3)	  &-0.00619(2)	  &-0.00471(2)  &$\cdots$\\
$d/r_s=1.0$ &-0.01392(6)	  &-0.0119(6)	  &-0.01064(5)	  &-0.00822(3)	  &-0.00613(2)  &-0.00419(1)\\
$d/r_s=1.3$ &-0.01811(5)	  &-0.015(1)	  &-0.0132(7)	  &-0.00995(5)	  &-0.00729(3)  &$\cdots$\\
$d/r_s=2.0$ &-0.02414(5)	  &-0.0191(2)	  &-0.0163(1)	  &-0.01189(7)	  &-0.00852(4)  &$\cdots$\\
$d/r_s=2.5$ &-0.02681(7)	  &-0.0206(2)	  &-0.0174(1)	  &-0.01252(8)	  &-0.0089(5)   &$\cdots$\\
$d/r_s=3.333$ &-0.02952(8)	  &-0.022(3)	  &-0.0184(2)	  &-0.0131(1)	  &-0.00921(6)  &$\cdots$\\
$d/r_s=5.0$ &-0.03259(8)	  &-0.0234(3)	  &-0.0193(2)	  &-0.0135(1)	  &-0.00948(6)  &-0.00617(3)\\
$d/r_s=7.0$ &-0.0345(1)	  &-0.024(3)	  &-0.0198(2)	  &-0.0137(1)	  &-0.0096(6)           &$\cdots$\\
$d/r_s=9.0$ &-0.0353(1)	  &-0.0243(3)	  &-0.0199(2)	  &-0.0138(1)	  &-0.00963(7)          &$\cdots$\\
\end{tabular}
\end{ruledtabular}
\end{table*}

\begin{table*}
\caption{\label{tab:partially_spin_polarized} Correlation energy per electron in partially spin-polarized systems. $\mu=r_s/d$ and $p=(N_{\uparrow}-N_{\downarrow})/N$.}
\begin{ruledtabular}
\begin{tabular}{ccccccc}
 & \multicolumn{3}{c}{$r_s = 1$} & \multicolumn{3}{c}{$r_s = 30$} \\
\cline{2-4} \cline{5-7}
 & $p=0.25$ & $p=0.5$ & $p=0.75$ & $p=0.25$ & $p=0.5$ & $p=0.75$ \\
\hline
$\mu=0.5$ &-0.0949(7) &-0.0818(5) &-0.0577(3) &-0.01137(4) &-0.01005(3) &-0.00776(2)\\
$\mu=2$ &-0.0502(1) &-0.04118(7) &-0.02583(6) &-0.006399(8) &-0.005415(7) &-0.003732(5)\\
\end{tabular}
\end{ruledtabular}
\end{table*}

\section{Treating the finite-size effect of the correlation energy}\label{appendix:FS}

\subsection{One-body size effect}
In metallic systems, the total energy exhibits a strong dependence on the boundary conditions applied to a finite simulation cell, due to the Fermi surface crossing between occupied and unoccupied states. This dependence on boundary conditions is a one-body effect, which is captured by Hartree-Fock (HF) calculations. It has also been shown that averaging over different twisted boundary conditions (TBC) can effectively reduce finite-size effects~\cite{Ceperley_PRE2001}. 

\begin{figure}
\includegraphics[width=1.0\textwidth]{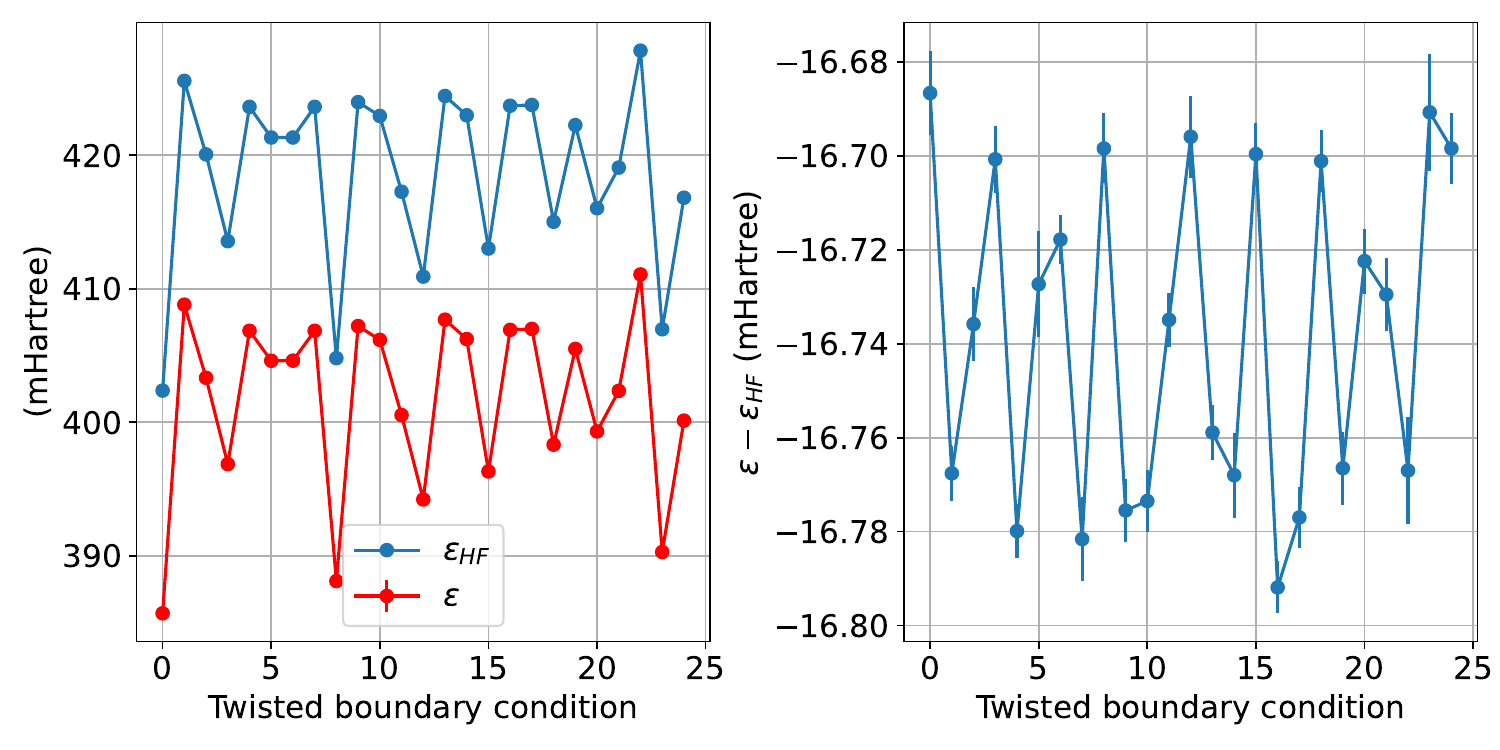}
\caption{Averaging the correlation energy across various twisted boundary conditions (TBC). The results presented in the plot correspond to a spin-unpolarized system with $r_s=1$ and $d=0.167$, and a total of $N_e=42$ electrons. The left panel illustrates how the total energy per electron $\epsilon$ and the Hartree-Fock (HF) energy per electron $\epsilon_{\mathrm{HF}}$ fluctuate similarly across different TBCs. In the right panel, it is evident that the fluctuation of $\epsilon-\epsilon_{\mathrm{HF}}$ is significantly reduced by a factor of $200$.}
\label{fig:twist}
\end{figure}

To compute the correlation energy, instead of twist-averaging the HF energy and total energy separately before subtracting them, we calculate the difference between the total and HF energies for each twisted boundary condition. We then average these differences to obtain the twist-averaged correlation energy at each finite size. This approach not only eliminates the one-body finite-size effect but also significantly reduces the error bars~\cite{Kwee_thesis2008}. Fig.~\ref{fig:twist} illustrates a typical example.

\subsection{Two-body size effect for Coulomb interaction without gate screening}
The finite-size effects arising from two-body interactions have been systematically addressed in recent studies~\cite{Chiesa_PRL2006, Drummond_PRB2008}. In the case of the HF energy, the only contribution to the finite-size effect from two-body terms comes from the potential energy:
\begin{equation}\label{eq:bc_potential}
     \frac{\Delta V}{N}\propto\frac{1}{A}\mathrm{lim}_{k\rightarrow0}v(k)S(k)=\frac{1}{A}\mathrm{lim}_{k\rightarrow0}\frac{S(k)}{k},
\end{equation}
where $A$ is the area of the simulation cell, $v(k)$ represents the Coulomb interaction in the momentum space, and $S(k)$ is the static structure factor. Within the HF framework, $S^{HF}(k)\propto k$ in 2D. Substituting this into Eq(\ref{eq:bc_potential}), one obtains:
\begin{equation}
    \Delta V_{HF}/N\propto \frac{1}{A}=\frac{n}{N},
\end{equation}
where $n$ is the density of the electrons.

For the quantum many-body ground state energy, the two-body finite-size effects arise from both the potential and kinetic energy terms. Under Coulomb interaction, the static structure factor behaves as $S(k)\propto k^{3/2}$. Substituting this into Eq(\ref{eq:bc_potential}), we find that the $N^{-1}$  term vanishes in the potential energy finite-size correction. On the other hand, the two-body finite-size correction to the kinetic energy is given by:
\begin{equation}\label{eq:kinetic_lead}
    \Delta T\propto \mathrm{lim}_{k\rightarrow0}k^2u(k),
\end{equation}
where $u(k)$ is the Jastrow factor.
$u(k)$ has the following relation with static structure factor within RPA\cite{Drummond_PRB2008}:
\begin{equation}\label{eq:jastrow}
    u(k)=-\frac{A}{N}\bigg\{\frac{-1}{2S_0(k)} +\bigg[\frac{1}{[2S_0(k)]^2}+\frac{N v(k)}{Ak^2}\bigg]^{1/2}\bigg\},
\end{equation}
where $S_0\propto k$ is the structure factor of the free electron gas in 2D. For small $k$, under Coulomb interaction, $v(k)=2\pi/k$, leading to $u(k)\propto k^{-3/2}$. Consequently, the $N^{-1}$ order also vanishes in finite-size correction in kinetic energy. 

The finite size correction of the correlation energy can be written as:
\begin{equation}
    \Delta E_c=\Delta V+\Delta T-\Delta V_{HF}.
\end{equation}
We can see that the leading order comes from the HF energy. Hence,
\begin{equation}
    \Delta E^{\mathrm{Coul}}_c/N\propto 1/N.
\end{equation}
Therefore, for systems with Coulomb interaction, we use the expression $\epsilon_c^{\mathrm{Coul}}(r_s,p,N)=a+b/N$ to fit the twist-averaged correlation energies obtained at different simulation cell sizes. The fitted $a$ is the correlation energy in the thermodynamic limit.

\subsection{Two-body size effect for dual-gate-screened Coulomb interaction}
In the presence of metallic gates, $v(k)=2\pi \mathrm{tanh}(kd)/k$. Consequently, the leading order of the finite-size scaling in the HF energy vanishes:
\begin{equation}\label{eq:tanh_potential}
     \frac{\Delta V}{N}\propto\frac{1}{A}\mathrm{lim}_{k\rightarrow0}\frac{S(k)\mathrm{tanh}(kd)}{k}=0.
\end{equation}
The finite size scaling of the total energy as well as the correlation energy hasn't been worked out yet. In this work, we use a quadratic polynomial $\epsilon_c(r_s,p,N)=a+b/N+c/N^2$ to fit our data. We computed three different system sizes ($N=42,90,122$) for each density and screening strength of the \sys. The error bars of the parameters $a,b,c$ are evaluated from the diagonal elements of the covariance matrix. 

\begin{figure}
\includegraphics[width=1.0\textwidth]{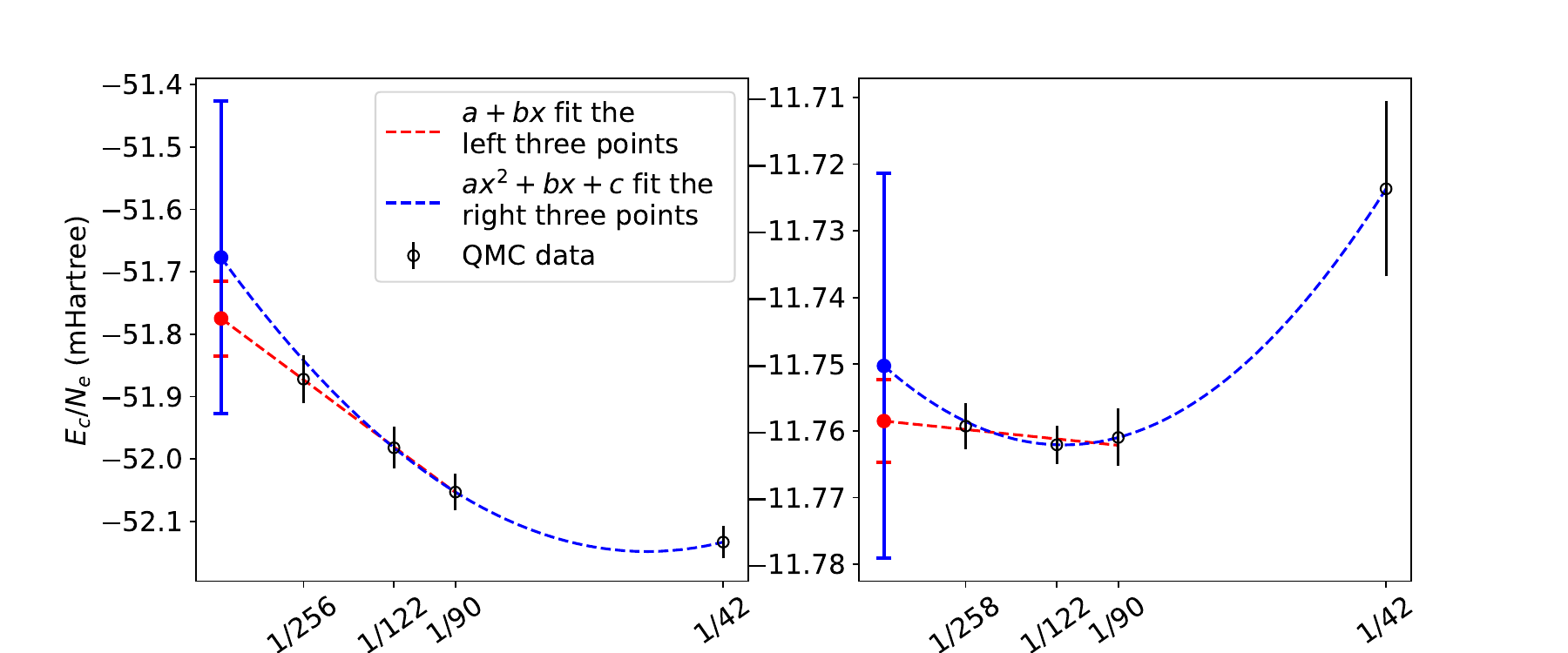}
\caption{Finite size extrapolation on the twist-averaged correlation energy of the spin-unpolarized \sys with gate screening. The left panel is from $r_s=1$ and $d=0.5$; the right panel is from $r_s=30$ and $d=60$. The results of the extrapolation are plotted on the left in each panel.}
\label{fig:extrapolate}
\end{figure}

To assess the quality of our finite-size extrapolation, we also computed larger system sizes for some parameters. In the two examples shown in Fig.~\ref{fig:extrapolate}, we can see that the extrapolated results from the quadratic polynomial fit to the three small system sizes are consistent with those from a linear fit on the three large system sizes, indicating the quadratic polynomial fit is appropriate. We applied this quadratic polynomial fit to all systems with gate screening.

\section{Perturbation analysis with the second quantization}\label{appendix:perturbation}
Second quantization is a convenient formalism for studying quantum many-body systems, particularly when dealing with systems of indistinguishable particles such as fermions or bosons. Our correlation energy at the high density limit ($r_s\rightarrow0$) and its asymptotic behaviors at the strongly screened limit are all obtained from perturbation analysis in the second quantization. Our derivations are described in this section.
\subsection{Preliminaries}
For completeness, we briefly summarize how the homogeneous electron gas (HEG) model is treated within the framework of the second quantization. This formalism is documented in various quantum many-body textbooks, for instance, Ref.~\cite{Fetter_book1971}.

The Hamiltonian of the two-dimensional electron gas immersed in a continuous and uniform background of positive charge under some form of two-body interaction $v(\abs{\mathbf{r}_i-\mathbf{r}_j})$ is:
\begin{equation}
\hat{H}=\hat{H}_{el}+\hat{H}_b+\hat{H}_{el-b},
\end{equation}
where 
\begin{equation}
    \hat{H}_{el}=\sum_{i=1}^N\frac{p_i^2}{2m} + \frac{1}{2}\sum_{i\ne j}^Nv(\abs{\mathbf{r}_i-\mathbf{r}_j})
\end{equation}
represents electron kinetic energy, and the electron-electron interactions
\begin{equation}
    \hat{H}_b=\frac{1}{2}\int\int d^2\mathbf{x}d^2\mathbf{x}'n(\mathbf{x})n(\mathbf{x'})v(\abs{\mathbf{x}-\mathbf{x'}})
\end{equation}
accounts for the background-background interactions, and
\begin{equation}
    \hat{H}_{el-b}=-\sum_{i=1}^N\int d^2\mathbf{x} n(\mathbf{x})v(\abs{\mathbf{x}-\mathbf{r}_i})
\end{equation}
describes the electron-background interactions. Making use of the translational invariance of the uniform positive-charge background, the $\hat{H}_b$ and $\hat{H}_{el-b}$ can be rewritten as:
\begin{equation}\label{eq:bg-bg}
    \hat{H}_{b}=\frac{1}{2}\frac{N^2}{A}\int d^2\mathbf{z}v(z),
\end{equation}
\begin{equation}\label{eq:el-bg}
    \hat{H}_{el-b=}-\frac{N^2}{A}\int d^2\mathbf{z} v(z).
\end{equation}
For the interaction part of $\hat{H}_{el}$, rewrite it in second quantization with plane-wave single particle orbitals $\psi_{k\sigma}(\mathbf{x})=A^{-1/2}e^{i\mathbf{k}\cdot\mathbf{x}}\eta_{\sigma}$~\cite{Fetter_book1971}, where $A$ is the area of the cell containing the electrons and $\eta_{\sigma}$ is the spin component. The matrix element of the two-body interaction is:
\begin{equation}
    \langle k_1\sigma_1k_2\sigma_2|v(z)|k_3\sigma_3k_4\sigma_4\rangle=\frac{1}{2A}\delta_{\sigma_1\sigma_3}\delta_{\sigma_2\sigma_4}\delta_{\mathbf{k}_1+\mathbf{k}_2,\mathbf{k}_3+\mathbf{k}_4}\int d^2\mathbf{z}e^{-i(\mathbf{k}_1-\mathbf{k}_3)\cdot\mathbf{z}}v(z).
\end{equation}
Next, change variables to make the conservation of momentum explicit:
\begin{equation}
\begin{split}
        &\mathbf{k}_1=\mathbf{k}+\mathbf{q} \,\,\,\, \mathbf{k}_3=\mathbf{k}\\
        &\mathbf{k}_2=\mathbf{p}-\mathbf{q} \,\,\,\, \mathbf{k}_4=\mathbf{p},
\end{split}
\end{equation}
and we obtain the interaction part of the Hamiltonian:
\begin{equation}
\begin{split}
\frac{1}{2A}\sum_{\sigma_1\sigma_2}\int d^2\mathbf{z}v(z) 
\left( \sum_{\mathbf{kpq}}{}' e^{-i\mathbf{q}\cdot\mathbf{z}} \hat{a}^{\dagger}_{\mathbf{k+q},\sigma_1}\hat{a}^{\dagger}_{\mathbf{p-q},\sigma_2} \hat{a}_{\mathbf{p},\sigma_2}\hat{a}_{\mathbf{k},\sigma_1} \right. 
\left. + \sum_{\mathbf{kp}}\hat{a}^{\dagger}_{\mathbf{k},\sigma_1}\hat{a}^{\dagger}_{\mathbf{p},\sigma_2} \hat{a}_{\mathbf{p},\sigma_2}\hat{a}_{\mathbf{k},\sigma_1}\right).
\end{split}
\end{equation}
Here, the first term in the above equation is the exchange term ($q\ne0$, as indicated by the prime on the sum symbol), and the second term is the direct term. 

Since the particle number is fixed in the system, the particle number operator is a number: $\sum_{\mathbf{p,\sigma_2}}\hat{a}^{\dagger}_{\mathbf{p},\sigma_2} \hat{a}_{\mathbf{p},\sigma_2}=N$. The direct term can then be rewritten as:
\begin{equation}
    \frac{1}{2A}\sum_{\sigma_1\sigma_2}\sum_{\mathbf{kp}}\int d^2\mathbf{z}v(z)(N^2-N).
\end{equation}
We can see that the first term above cancels the sum of Eq.~(\ref{eq:bg-bg}) and Eq.~(\ref{eq:el-bg}). The second term, when divided by the total number of electrons $N$, vanishes under the thermodynamic limit. Similarly, we can obtain the matrix element of the kinetic operator under the plane wave basis:
\begin{equation}
    \hat{T}=\sum_{k\sigma}\frac{k^2}{2}a^{\dagger}_{k\sigma}a_{k\sigma}.
\end{equation}

Therefore, the Hamiltonian under the second quantization is:
\begin{equation}\label{eq:ac}
    \hat{H}= \sum_{k\sigma}\frac{k^2}{2}a^{\dagger}_{k\sigma}a_{k\sigma} + \frac{1}{2A}\sum_{\sigma_1\sigma_2}\sum_{\mathbf{kpq}}{}'v(\mathbf{q}) \hat{a}^{\dagger}_{\mathbf{k+q},\sigma_1}\hat{a}^{\dagger}_{\mathbf{p-q},\sigma_2} \hat{a}_{\mathbf{p},\sigma_2}\hat{a}_{\mathbf{k},\sigma_1},
\end{equation}
where 
\begin{equation}
    v(\mathbf{q})=\int d^2\mathbf{z}v(z)  e^{-i\mathbf{q}\cdot\mathbf{z}}.
\end{equation}
We denote the kinetic term by $\hat{H}_0$ and the interaction term by $\hat{H}_1$.

Under zero temperature and without interaction, electrons occupy the energy levels following a heavyside step function. The electron density uniquely determines the Fermi vector $k_F$. This can be seen by evaluating the total number of electrons in area $A$:
\begin{equation}\label{eq:N}
N=\sum_{\mathbf{k}\sigma}\langle\Phi_0|\hat{n}_{\mathbf{k}\sigma}|\Phi_0\rangle=\sum_{\mathbf{k}\sigma}\theta(k_F-\mathbf{k}),
\end{equation}
where 
\begin{equation}\label{eq:heavy-side-step}
    \theta(x)=
\begin{cases}
    1, \;\;\;x>0 & \\
    0, \;\;\;x<0 &.
\end{cases}
\end{equation}
Imagine that the area is a square cell with side length $L$ and periodic boundary condition. The electron momentum is then discrete
\begin{equation}
    k_{x (y)}=\frac{2\pi n}{L} \;\;\;\; n=1,2,3,...
\end{equation}
with equal spacing
\begin{equation}
    \Delta k_{x (y)}=\frac{2\pi}{L}.
\end{equation}
The discrete sum over a momentum $\mathbf{k}$ can then be rewritten as:
\begin{equation}
    \sum_{\mathbf{k}}=\sum_{k_x}\left(\Delta k_x \frac{L}{2\pi}\right) \sum_{k_y}\left(\Delta k_y \frac{L}{2\pi}\right).
\end{equation}
As $L\rightarrow\infty$, $\Delta k\rightarrow0$ and the discrete sum in Eq.~(\ref{eq:N}) becomes an integral:
\begin{equation}\label{eq:sum2Integral}
N=\sum_{k\sigma}\theta(k_F-\mathbf{k})=A(2\pi)^{-2}\sum_{\sigma}\int d^2\mathbf{k}\theta(k_F-\mathbf{k})=\frac{1}{4\pi}Ak_F^2\,2,
\end{equation}
where the factor $2$ comes from the spin sum. Note that for spin polarized case, there is no spin sum any more in Eq.~(\ref{eq:N}). Therefore, the number of electrons in area $A$ in the spin polarized case:
\begin{equation}\label{eq:V_kF_spin_polarized}
    N=\frac{1}{4\pi}Ak_F^2.
\end{equation}
Note that $A/N=\pi r_s^2$. Therefore,
\begin{equation}\label{eq:k_F}
    k_F=
\begin{cases}
    \frac{\sqrt{2}}{r_s}, \;\;\;p=0 & \\
    \frac{2}{r_s}, \;\;\;p=1. &
\end{cases}
\end{equation}

\subsection{Perturbation analysis}
When electron density is high or the screening is strong, $\hat{H}_1$ is weak compared to $\hat{H}_0$ and we can start from the non-interacting ground state and compute the corrections by perturbation analysis. This can be carried out conveniently from Goldstone's theorem~\cite{Goldstone_1957PRSLA,Fetter_book1971}.
\begin{equation}\label{eq:Goldstone}
    E-E_0=\langle\Phi_0|\hat{H}_1\sum_{n=1}^{\infty}\left(\frac{1}{E_0-\hat{H}_0}\hat{H}_1\right)^{n-1} |\Phi_0\rangle_{\mathrm{connected}},
\end{equation}
where $E$ is the many-body ground state energy, $E_0$ the eigenvalue of the non-interacting Hamiltonian $\hat{H}_0$, and $\ket{\Phi_0}$ the eigenstate of $\hat{H}_0$. $(E_0-\hat{H}_0)^{-1}$ propagates the state each time the state is acted upon by an interaction $\hat{H}_1$. The subscript ``connected'' means that only the connected diagrams are included in the sum. This is a restatement of the time-independent perturbation which is also known as the Rayleigh-Schrodinger perturbation. 

For $n=1$, the first-order correction to the non-interacting energy is
\begin{equation}\label{eq:Ex}
    E^{(1)}=\bra{\Phi_0}\hat{H}_1\ket{\Phi_0}_{\mathrm{connected}}=\frac{1}{2A}\sum_{\mathbf{k\,p\,q}}'\sum_{\sigma_1\sigma_2}v(\mathbf{q})\bra{\Phi_0}\hat{a}^{\dagger}_{\mathbf{k+q},\sigma_1}\hat{a}^{\dagger}_{\mathbf{p-q},\sigma_2}\hat{a}_{\mathbf{p},\sigma_2}\hat{a}_{\mathbf{k},\sigma_1}\ket{\Phi_0}.
\end{equation}
From now on we use $v(\mathbf{q})$ to represent the interaction in Eq.~(\ref{eq:kspace}) for simplicity. As defined previously, the prime on the sum symbol means: omit the term $\mathbf{q}=0$. There is only one way of pairing the creation and annihilation operators that leads to the non-vanishing result:
\begin{equation}
\begin{cases}
    \mathbf{k+q},\sigma_1 = \mathbf{p},\sigma_2 &   \\
    \mathbf{p-q},\sigma_2 = \mathbf{k},\sigma_1  & .
\end{cases}
\end{equation}
Eq.~(\ref{eq:Ex}) can then be rewritten as:
\begin{equation}
\begin{split}
\frac{1}{2A}\sum_{\mathbf{k\,q}}'\sum_{\sigma_1}v(\mathbf{q})\bra{\Phi_0}\hat{a}^{\dagger}_{\mathbf{k+q},\sigma_1}\hat{a}^{\dagger}_{\mathbf{k},\sigma_1}\hat{a}_{\mathbf{k+q},\sigma_1}\hat{a}_{\mathbf{k},\sigma_1}\ket{\Phi_0}=
-\frac{1}{2A}\sum_{\mathbf{k\,q}}'\sum_{\sigma_1}v(\mathbf{q})\bra{\Phi_0}\hat{n}_{\mathbf{k+q},\sigma_1}\hat{n}_{\mathbf{k},\sigma_1}\ket{\Phi_0}.
\end{split}
\end{equation}
Similar to Eq.~(\ref{eq:N}), the above equation can be rewritten as:
\begin{equation}\label{eq:general_Ex}
    -\frac{1}{2A}\sum_{\mathbf{k\,q}}'\sum_{\sigma_1}v(\mathbf{q})\theta\left(k_F-\mathbf{\abs{k+q}}\right)\theta\left(k_F-\mathbf{\abs{k}}\right).
\end{equation}
Converting the sums to integrals under thermodynamic limit as in Eq.~(\ref{eq:sum2Integral}), we obtain:
\begin{equation}\label{eq:2dEx}
    E^{(1)}=- \frac{1}{2A}\frac{A^2}{(2\pi)^4}2\int d^2\mathbf{q}v(\mathbf{q})\int d^2\mathbf{k}\theta\left(k_F-\abs{\mathbf{k+q}}\right)\theta\left(k_F-\abs{\mathbf{k}}\right),
\end{equation}
where again $A=L^2$, and the extra factor of $2$ comes from the sum over spin index $\sigma_1$. Evaluating this integral~\cite{Fetter_book1971}, we obtain the exchange energy per electron:
\begin{equation}\label{eq:exchange}
    \epsilon_{x}(r_s,\mu,p)=\frac{-\sqrt{2}}{\pi r_s} \int_0^1 dx \sum_{s=+,-} P_s^3 \mathrm{tanh}\left(2\sqrt{2}P_s x/\mu\right)
    \left(\arccos{x} - x\sqrt{1-x^2}\right),
\end{equation}
where $P_{\pm} = \sqrt{1 \pm p}$ and $p$ is the polarization ratio, as was defined in Eq.~(\ref{eq:bc_HF}) in the main text.

For $n=2$, the second-order correction

\begin{equation}\label{eq:second-order-perturbation}
    E^{(2)}=\langle\Phi_0|\hat{H}_1\left(\frac{1}{E_0-\hat{H}_0}\hat{H}_1\right) |\Phi_0\rangle_{\mathrm{connected}}=\sum_{n\neq0}\frac{\langle\Phi_0|\hat{H}_1|n\rangle\langle n|\hat{H}_1|\Phi_0\rangle}{E_0-E_n}.
\end{equation}

Substituting the interacting part of the Hamiltonian $\hat{H}_1$ into the numerator of Eq.~(\ref{eq:second-order-perturbation}), we obtain
\begin{equation}\label{eq:second-order-perturbation-substitute-H1}
    \frac{1}{(2A)^2}\sum_{\mathbf{k\,p\,q}}'\sum_{\sigma_1\sigma_2} \sum_{\mathbf{k'\,p'\,q'}}'\sum_{\sigma_1'\sigma_2'}v(\mathbf{q})v(\mathbf{q'})\bra{\Phi_0}\hat{a}^{\dagger}_{\mathbf{k+q},\sigma_1}\hat{a}^{\dagger}_{\mathbf{p-q},\sigma_2} \hat{a}_{\mathbf{p}',\sigma_2'}\hat{a}_{\mathbf{k}',\sigma_1'}\ket{n}\bra{n}\hat{a}^{\dagger}_{\mathbf{k'+q'},\sigma_1'}\hat{a}^{\dagger}_{\mathbf{p'-q'},\sigma_2'} \hat{a}_{\mathbf{p}',\sigma_2'}\hat{a}_{\mathbf{k}',\sigma_1'}\ket{\Phi_0}.
\end{equation}
The final factor $\bra{n}\hat{a}^{\dagger}_{\mathbf{k'+q'},\sigma_1'}\hat{a}^{\dagger}_{\mathbf{p'-q'},\sigma_2'} \hat{a}_{\mathbf{p}',\sigma_2'}\hat{a}_{\mathbf{k}',\sigma_1'}\ket{\Phi_0}$ in Eq.~(\ref{eq:second-order-perturbation-substitute-H1}) equals $1$ when $\ket{n}=\hat{a}^{\dagger}_{\mathbf{k'+q'},\sigma_1'}\hat{a}^{\dagger}_{\mathbf{p'-q'},\sigma_2'} \hat{a}_{\mathbf{p}',\sigma_2'}\hat{a}_{\mathbf{k}',\sigma_1'}\ket{\Phi_0}$ but otherwise vanishes. This is due to the orthogonality of the plane wave orbitals in the non-interacting state. Therefore, Eq.~(\ref{eq:second-order-perturbation-substitute-H1}) reduces to
\begin{equation}
     \frac{1}{(2A)^2}\sum_{\mathbf{k\,p\,q}}'\sum_{\sigma_1\sigma_2} \sum_{\mathbf{k'\,p'\,q'}}'\sum_{\sigma_1'\sigma_2'}v(\mathbf{q})v(\mathbf{q'})\bra{\Phi_0}\hat{a}^{\dagger}_{\mathbf{k+q},\sigma_1}\hat{a}^{\dagger}_{\mathbf{p-q},\sigma_2} \hat{a}_{\mathbf{p}',\sigma_2'}\hat{a}_{\mathbf{k}',\sigma_1'}
   \hat{a}^{\dagger}_{\mathbf{k'+q'},\sigma_1'}\hat{a}^{\dagger}_{\mathbf{p'-q'},\sigma_2'} \hat{a}_{\mathbf{p}',\sigma_2'}\hat{a}_{\mathbf{k}',\sigma_1'}\ket{\Phi_0}.
\end{equation}
Now we can apply Wick's theorem and sum all contractions. This $n=2$ term corresponds to the scattering of two electrons. The two electrons are excited to unoccupied states and then fall back to either their original states or their each other's original states~\cite{Pines_book1999,Fetter_book1971}. The denominator in Eq.~(\ref{eq:second-order-perturbation}) is the energy difference between the non-interacting ground state and the state with two-electron excitation from the ground state:
\begin{equation}
    \frac{(\mathbf{p+q})^2}{2}+\frac{(\mathbf{-k-q})^2}{2}-\frac{p^2}{2}-\frac{k^2}{2}.
\end{equation}

Similar to what we did in calculating the first-order correction, we obtain the second-order correction
\begin{equation}\label{eq:2nd_pert}
    \begin{split}
        \epsilon_c^{(2)}(\mu)=\frac{k_F^2}{8\pi^3}\int d^2\mathbf{q} v(\abs{\mathbf{q}}k_F)\int_{\abs{\mathbf{p+q}}>1} d^2\mathbf{p}\int_{\abs{\mathbf{k+q}}>1} d^2\mathbf{k}
    \frac{\theta(1-\mathbf{p})\theta(1-\mathbf{k})}{\mathbf{q}^2+\mathbf{q\cdot (p+k})}\bigg[v(\abs{\mathbf{q+ p + k}}k_F)-
    a v\left(\abs{\mathbf{q}}k_F\right)\bigg].
    \end{split}
\end{equation}
Note that $\mu=r_s/d$ and $k_F\propto r_s^{-1}$. Here $\mathbf{p,k,q}$ are all in the unit of Fermi vector $k_F$. $\theta(x)$ is the heavyside step function defined in Eq.~(\ref{eq:heavy-side-step}). 

\begin{figure}
\includegraphics[scale=0.2]{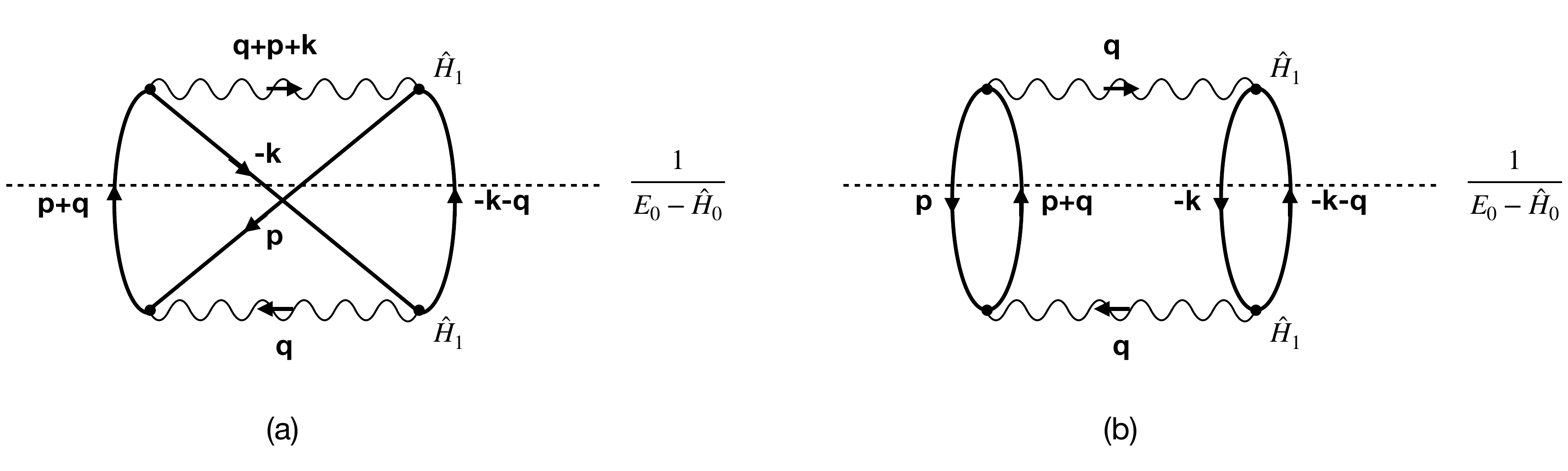}
\caption{The exchange (a) and direct (b) term represented by Goldstone diagrams.}
\label{fig:diagram}
\end{figure}

The first term in Eq.~(\ref{eq:2nd_pert}) is the exchange term and the second term is the direct term. These two terms can be represented by the Goldstone diagrams shown in Fig.~\ref{fig:diagram}(a) and Fig.~\ref{fig:diagram}(b), respectively.
We point out that Eq.~(\ref{eq:2nd_pert}) is a function of $\mu$ and doesn't explicitly depend on $r_s$ ($v(\abs{\mathbf{q}}k_F)\propto k_F^{-1}$). In addition, it is guaranteed to be finite because even in the absence of screening the correlation energy of the second-order perturbation does not diverge~\cite{Rajagopal_PRB1977}.

\begin{figure}
\centering
\includegraphics[scale=0.3]{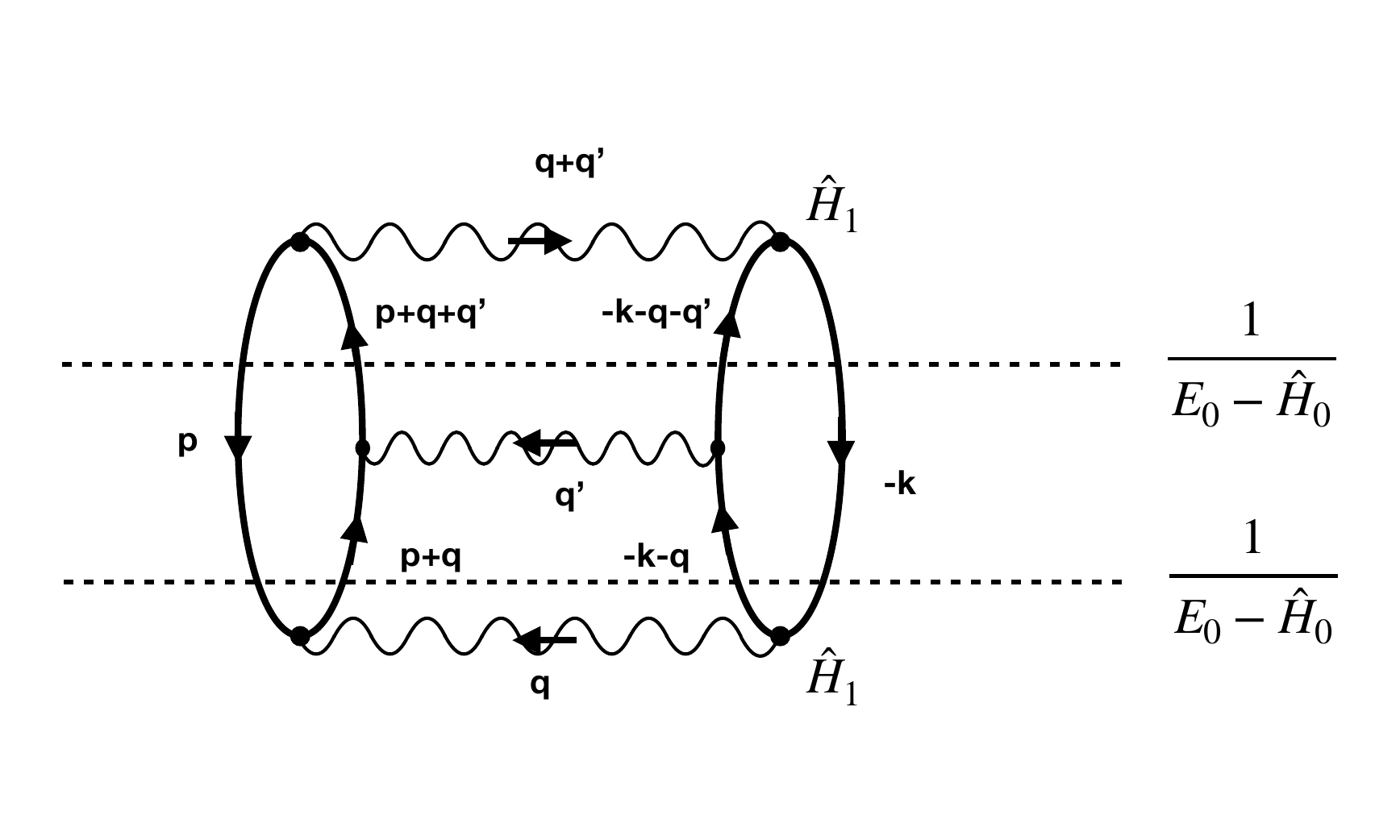}
\caption{A diagram in the third order perturbation.}
\label{fig:diagram_3rd}
\end{figure}

For $n=3$, the third-order perturbation can be conveniently represented with Goldstone diagrams. Since all diagrams in the same order should give the same asymptotic limit, we study a single diagram in the third order, as shown in Fig.~\ref{fig:diagram_3rd}.
From this diagram, we can see the third-order correction is proportional to 
\begin{equation}\label{eq:3rd_order}
\begin{split}
    &\frac{1}{(2A)^3}\frac{A^4}{(2\pi)^8}k_F^4\int d^2\mathbf{k}\int d^2\mathbf{p}\int d^2\mathbf{q}\int d^2\mathbf{q'}
    v(\abs{\mathbf{q}}k_F)v(\abs{\mathbf{q}'}k_F)v(\abs{\mathbf{q+q'}}k_F)\\
    &\frac{\theta(1-\mathbf{p})\left[1-\theta(1-\mathbf{p}-\mathbf{q})\right]
    \theta(1-\mathbf{k})\left[1-\theta(1-\mathbf{k}-\mathbf{q})\right]
    \left[1-\theta(1-\mathbf{p}-\mathbf{q}-\mathbf{q'})\right]
    \left[1-\theta(1-\mathbf{k}-\mathbf{q}-\mathbf{q'})\right]}{\mathbf{q}\cdot(\mathbf{q+p+k})\;\;\mathbf{q'}\cdot(\mathbf{2q+p+k+q'})},
\end{split}
\end{equation}
where again all the momentum are in the unit of $k_F$.

\subsection{Strongly screened limit}
\subsubsection{Power counting}
From the previous analysis, we find that the $n$th term is proportional to: 
\begin{equation}
    A^{n+1}\frac{1}{A^{n}}k_F^{2(n+1)}\int d\mathbf{p}\int d\mathbf{k}\left[\int d\mathbf{q}\right]^{n-1}\frac{f(\mathbf{p}, \mathbf{k}, \mathbf{q}_1,\cdots, \mathbf{q}_{n-1})}{k_F^{2(n-1)}}v^n,
\end{equation}
where $f(\mathbf{p}, \mathbf{k},\mathbf{q}_1,\cdots, \mathbf{q}_{n-1})$ is the constraint that the momentum involved in the process of electron interactions must satisfy, for example the $\theta$ functions in Eq.~(\ref{eq:2dEx}). $A$ is again the area of the cell that contains the electrons. Note that $A=N\pi r_s^2$, where $N$ denotes the total number of electrons. The momenta $\mathbf{p,k}$, and $\mathbf{q}_i$ here are all in units of the Fermi wave vector $k_F$. 

The interaction
\begin{equation}
    v\propto \frac{\mathrm{tanh}(\{\mathbf{p,k,q_1,q_2,...,q_{n-1}}\}\mu^{-1})}{\{\mathbf{p,k,q_1,q_2,...,q_{n-1}}\}k_F},
\end{equation}
where $\{\mathbf{p,k,q_1,q_2,...,q_{n-1}}\}$ represents different linear combination of all the momentum involved. The thermodynamic limit is taken by letting both $N\rightarrow\infty$ and $A\rightarrow\infty$, while keeping $N/A$ fixed. Noting that $k_F\propto r_s^{-1}$, we find that the $n$th term is proportional to 
\begin{equation}\label{eq:nth_order}
    r_s^{n-2}\int d\mathbf{p}\int d\mathbf{k}\left[\int d\mathbf{q}\right]^{n-1} \left(\frac{\mathrm{tanh}(\{\mathbf{p,k,q_1,q_2,...,q_{n-1}}\}\mu^{-1})}{\{\mathbf{p,k,q_1,q_2,...,q_{n-1}}\}}\right)^n.
\end{equation}

In the strongly screened limit of the Fermi gas, the electron-electron interaction becomes short-ranged due to effective screening of the long-range Coulomb potential. The physics of the dilute, short-range interacting Fermi gas has been studied extensively using perturbation theory, expanding in powers of the electron density~\cite{Engelbrecht_PRB1992}. In dual-gate devices, the screened Coulomb interaction strength, scaling as $1/\mu$, suppresses the logarithmic divergence caused by the particle-particle fluctuations, enabling a perturbative treatment of the interaction effects. We expect that the $n$-th order term in the perturbation series exhibits the scaling behavior:
\begin{equation}    
\int d\mathbf{p}\int d\mathbf{k}\left[\int d\mathbf{q}\right]^{n-1}\left(\frac{\mathrm{tanh}(\{\mathbf{p,k,q_1,q_2,...,q_{n-1}}\}\mu^{-1})}{\{\mathbf{p,k,q_1,q_2,...,q_{n-1}}\}}\right)^n\propto\mu^{-n},
\end{equation}
To conclude, the $n$th order contribution scales as
\begin{equation}\label{eq:n-th-order-asym}
 r_s^{n-2}\mu^{-n}    
\end{equation}
This power counting analysis is accurate up to logarithmic corrections. In the following two subsections, we further analyze the structure of these logarithmic corrections.

\subsubsection{Spin-unpolarized case}
The leading order of the asymptotic behavior in the strongly screened limit comes from the second-order correction shown in Eq.~(\ref{eq:2nd_pert}). As $\mu\rightarrow\infty$, following Zecca et. al~\cite{Zecca_PRB2004}, we work in the polar coordinates and split the radial integration over $q$ ($q=\abs{\mathbf{q}}$) into two parts
\begin{equation}\label{eq:split}
    \int_0^{\infty}dq=\int_0^{q_1}dq+\int_{q_1}^{\infty}dq,
\end{equation}
where $1\ll q_1\ll \mu$. 

In the first part, since $q \in [0, q_1]$, we have $qk_Fd \rightarrow 0$ as $\mu = r_s/d \rightarrow \infty$. Similarly, because $\mathbf{p}$ and $\mathbf{k}$ lie below the Fermi surface, we also have $\abs{\mathbf{q+p+k}}k_Fd \rightarrow 0$. Expanding the $\mathrm{tanh}$ functions in Eq.~(\ref{eq:2nd_pert}) into Taylor series. we obtain 
\begin{equation}\label{eq:2nd_pert_Taylor}
    \begin{split}
        {}^1\epsilon_c^{(2)}(\mu)=&\frac{k_F^2}{8\pi^3}\int_{0<q<q_1} d^2\mathbf{q} \int_{\abs{\mathbf{p+q}}>1} d^2\mathbf{p}\int_{\abs{\mathbf{k+q}}>1} d^2\mathbf{k}
    \frac{\theta(1-\mathbf{p})\theta(1-\mathbf{k})}{\mathbf{q}^2+\mathbf{q\cdot (p+k})}\\
    &\bigg[
    \frac{\abs{\mathbf{q+ p + k}}k_Fd-\frac{1}{3}(\abs{\mathbf{q+ p + k}}k_Fd)^3+O\left((\abs{\mathbf{q+ p + k}}k_Fd\right)^5)}{\abs{\mathbf{q+ p + k}}k_F}
    \frac{qk_Fd-\frac{1}{3}(qk_Fd)^3+O\left(qk_Fd\right)^5}{qk_F}
    -\\
    &a\left(\frac{qk_Fd-\frac{1}{3}(qk_Fd)^3+O\left(qk_Fd\right)^5}{qk_F}\right)^2\bigg].
    \end{split}
\end{equation}
Taking the leading order terms, we obtain the asymptotic limit of the first part:
\begin{equation}\label{eq:1stPart_2nd_pert}
    \begin{split}
        {}^1\epsilon_c^{(2)}(\mu)=\frac{1-
    a}{2a\pi^3\mu^2}\int_{0<q<q_1} d^2\mathbf{q} \int_{\abs{\mathbf{p+q}}>1} d^2\mathbf{p}\int_{\abs{\mathbf{k+q}}>1} d^2\mathbf{k}\frac{\theta(1-\mathbf{p})\theta(1-\mathbf{k})}{\mathbf{q}^2+\mathbf{q\cdot (p+k})}=O\left(\frac{1}{\mu^2}\right),
    \end{split}
\end{equation}
where we have used the definition $\mu\equiv r_s/d$ and Eq.~(\ref{eq:k_F}). Here
\begin{equation}\label{eq:a}
    a=
\begin{cases}
    1, \;\mathrm{spin \;polarized} & \\
    2, \;\mathrm{spin \;unpolarized}.
\end{cases}
\end{equation}

In the second part, $q\in[q_1,\infty)$. Now the arguments in $\mathrm{tanh}$ are no longer small. In this case, since $\abs{\mathbf{k}}<1$ and $\abs{\mathbf{p}}<1$, we have
\begin{equation}
    \abs{\mathbf{q+p+k}}\approx q.
\end{equation}
Substitute it into Eq.~(\ref{eq:2nd_pert}), and note that when $q \geq q_1 \gg 1$, the constraints $\abs{\mathbf{p+q}} > 1$ and $\abs{\mathbf{k+q}} > 1$ are automatically satisfied. In this case, the integrals over $\mathbf{p}$ and $\mathbf{k}$ can be performed independently, and each of them equals $1$. This yields:
\begin{equation}\label{eq:pre_2ndPart_2nd_pert}
    \begin{split}
        {}^2\epsilon_c^{(2)}(\mu)&=\frac{1-
    a}{4}\int_{q_1}^{\infty} dq\frac{\mathrm{tanh}^2(qk_Fd)}{q^3}. \\
    &=\frac{1-
    a}{4}\int_{q_1}^{1/k_Fd} dq\frac{(k_Fd)^2}{q} + O\left(\frac{1}{\mu^2}\right) \\
    &=-\frac{1-
    a}{4}\mathrm{ln}(k_Fd)(k_Fd)^2+ O\left(\frac{1}{\mu^2}\right) \\
    &=\frac{1-
    a}{2}\frac{\mathrm{ln} \mu}{\mu^2}+ O\left(\frac{1}{\mu^2}\right), \\
    \end{split}
\end{equation}
where we have used the definition $\mu\equiv r_s/d$ and Eq.~(\ref{eq:k_F}).

Combining with Eq.~(\ref{eq:1stPart_2nd_pert}), and noting that $a=2$ in the spin-unpolarized case, the leading order is then:
\begin{equation}
    \epsilon_c^{(2)}(\mu)= {}^1\epsilon_c^{(2)}(\mu)+ {}^2\epsilon_c^{(2)}(\mu)=-0.5\frac{\mathrm{ln}\mu}{\mu^2}+O\left(\frac{1}{\mu^2}\right).
\end{equation}

\subsubsection{Spin-polarized case}
In the spin-polarized case, which corresponds to $a=1$, both Eq.~(\ref{eq:1stPart_2nd_pert}) and Eq.~(\ref{eq:pre_2ndPart_2nd_pert}) vanish. The leading order in the asymptotic limit comes from the third-order perturbation.

To derive the asymptotic limit from the third-order perturbation shown in Eq.~(\ref{eq:3rd_order}), we decompose the double integral over an infinite domain into four separate double integrals with finite and infinite limits,
\begin{equation}\label{eq:terms}
    \int_0^{\infty}dq\int_0^{\infty}dq'=\int_0^{q_1}dq\int_0^{q_1}dq'+\int_0^{q_1}dq\int_{q_1}^{\infty}dq'+\int_0^{q_1}dq'\int_{q_1}^{\infty}dq+\int_{q_1}^{\infty}dq\int_{q_1}^{\infty}dq',
\end{equation}
For the first term, expanding the potentials $v(\abs{\mathbf{q}}k_F)$, $v(\abs{\mathbf{q}'}k_F)$, and $v(\abs{\mathbf{q+q'}}k_F)$ in Eq.~(\ref{eq:3rd_order}) in Taylor series and retaining leading order terms yields:
\begin{equation}
\begin{split}
    \frac{1}{k_F}\int d^2\mathbf{k}\int d^2\mathbf{p}\int d^2\mathbf{q}\int d^2\mathbf{q'}
    \frac{(k_Fd)^3f(\mathbf{k,p,q,q'})}{\mathbf{q}\cdot(\mathbf{q+p+k})\;\;\mathbf{q'}\cdot(\mathbf{2q+p+k+q'})}=O\left(\frac{r_s}{\mu^3}\right),
\end{split}
\end{equation}
where we again have used the definition $\mu\equiv r_s/d$ and Eq.~(\ref{eq:k_F}) and $f(\mathbf{k,p,q,q'})$ represents the product of all $\theta$ functions in the numerator in Eq.~(\ref{eq:3rd_order}) for simplicity. 

The second and third terms in Eq.~(\ref{eq:terms}) are the same, so we evaluate the second term only. The second term is small compared to the first term because in the denominator $q'\in[q_1,\infty)$. Since $1\ll q_1 < q'$, $\theta(1-\mathbf{p}-\mathbf{q}-\mathbf{q'})=\theta(1-\mathbf{k}-\mathbf{q}-\mathbf{q'})=0$ and $\mathbf{2q+p+k+q'}\approx\mathbf{q'}$. Substituting them back and expanding $v(\abs{\mathbf{q}}k_F)$ into Taylor series, Eq.~(\ref{eq:3rd_order}) can be simplified to be
\begin{equation}\label{eq:3rd_order_2}
\begin{split}
    2\pi d\int_{q_1}^{\infty} dq'\frac{\mathrm{tanh}^2(q'dk_F)}{q'^4}\int d^2\mathbf{k}\int d^2\mathbf{p}\int_0^{q_1} dq\int_0^{2\pi}d\theta_q
    \frac{\theta(1-\mathbf{p})\left[1-\theta(1-\mathbf{p}-\mathbf{q})\right]
    \theta(1-\mathbf{k})\left[1-\theta(1-\mathbf{k}-\mathbf{q})\right]}{\mathbf{q}\cdot(\mathbf{q+p+k})\;\;}
\end{split}
\end{equation}
Splitting the integral over $q'$ similar to what we did in Eq.~(\ref{eq:pre_2ndPart_2nd_pert}), we obtain 
\begin{equation}\label{eq:3rd_order_2_asym}
     (k_Fd)^2 + O\big( (k_Fd)^3 \big),
\end{equation}
Substituting Eq.~(\ref{eq:3rd_order_2_asym}) back to Eq.~(\ref{eq:3rd_order_2}) and using Eq.~(\ref{eq:k_F}) and the definition $\mu\equiv r_s/d$, we find the asymptotic behavior to be: $O\left(\frac{r_s}{\mu^3}\right)$. 

The fourth term in Eq.~(\ref{eq:terms}) is small compared to the second and third terms because in the denominator both $q'\in[q_1,\infty)$ and $q\in[q_1,\infty)$. The integrand is less singular than the small $q$ and $q'$ limit. We expect the asymptotic behavior to be $O\left(\frac{r_s}{\mu^3}\right)$.
We validated it numerically by fitting the results from the integral. Therefore, 
\begin{equation}
    \epsilon_c^{(3)}(\mu)=O\left(\frac{r_s}{\mu^3}\right)
\end{equation}

\subsection{Infinite density limit}
In the infinite density limit, where $r_s \rightarrow 0$ while keeping the screening strength $\mu$ finite, the scaling behavior differs from that of the strongly screened limit discussed in the previous subsections. According to Eq.~(\ref{eq:nth_order}), terms corresponding to $n>2$ vanish at $r_s=0$. Consequently, the $n=2$ term becomes the exact correlation energy at $r_s=0$. This key observation is utilized in the main text to analyze the system's behavior in the infinite density limit. 

Furthermore, when taking the infinite density limit ($r_s \rightarrow 0$) first and then considering the strongly screened limit, power counting suggests that the correlation energy scales as $1/\mu^2$. However, a more thorough analysis, similar to that in the previous subsections, reveals distinct behaviors for the spin-polarized and spin-unpolarized cases.

In the spin-unpolarized case, the naive power counting holds, and the correlation energy indeed scales as $1/\mu^2$. Conversely, for the spin-polarized case, the prefactor of the $1/\mu^2$ term vanishes exactly. The next-leading-order correction arises from from keeping the higher-order term in one of the $v(q)$ in the second-order perturbation result, i.e., Eq.~(\ref{eq:2nd_pert_Taylor}). This correction is of the order $1/\mu^4$.

The resulting scaling of the second-order correlation energy $\epsilon_c^{(2)}(\mu)$ can be summarized as follows:

\begin{equation}\label{eq:a}
    \epsilon_c^{(2)}(\mu)=
\begin{cases}
    O\left(\frac{1}{\mu^4}\right)+O\left(\frac{r_s}{\mu^5}\right), \;\mathrm{spin \;polarized} & \\
    -0.5\frac{\mathrm{ln(\mu)}}{\mu^2}+O\left(\frac{1}{\mu^2}\right), \;\mathrm{spin \;unpolarized}.
\end{cases}
\end{equation}

\section{Hartree Fock energy}\label{appendix:HF}
The Hartree-Fock (HF) energy is composed of electrostatic (Hartree) energy, exchange (Fock) energy, and kinetic energy. The Hartree energy cancels out due to the presence of the background. The exchange energy, owing to translational invariance, is equivalent to the first-order perturbation energy with respect to the electron-electron interaction~\cite{Fetter_book1971, Pines_book1999}. Thus, the HF energy can be expressed as:
\begin{equation}\label{eq:HF}
    \epsilon_{\mathrm{HF}}=\frac{1}{2r_s^2}\frac{(1+p)^2+(1-p)^2}{2}+\epsilon_{x}(r_s,\mu,p),
\end{equation}
where the first term is the kinetic energy, and the second term is the exchange energy shown in Eq.~(\ref{eq:exchange}). It can be rewritten as :
\begin{equation}
    \epsilon_{x}(r_s,\mu,p=0)=\epsilon_{x}^{\mathrm{Coul}}(r_s,p)F(\mu,p),
\end{equation}
where 
\begin{equation}\label{eq:F}
    F(\mu,p)=\frac{3}{4}\int_0^1 dx \sum_{s=+,-}P_s^3\mathrm{tanh}\left(\frac{2\sqrt{2}P_sx}{\mu}\right)\left(\arccos{x} - x\sqrt{1-x^2}\right).
\end{equation}

To conveniently implement the exchange energy to a computer routine, we use a Padé form to fit it. Let $Q_{\pm}=P_{\pm}/\mu$ and the two terms in Eq.~(\ref{eq:exchange}) can be rewritten as:
\begin{equation}
    {}^{\pm}\epsilon_{x}(r_s,\mu,p)=\frac{C_x}{r_s} \left(\mu Q_{\pm}\right)^3\int_0^1 dx  \;\mathrm{tanh}\left(2\sqrt{2}Q_{\pm} x\right)
    \left(\arccos{x} - x\sqrt{1-x^2}\right).
\end{equation}
Now we fit this one-dimensional function
\begin{equation}
    f(Q)=\int_0^1 dx \; \mathrm{tanh}\left(2\sqrt{2}Q x\right)
    \left(\arccos{x} - x\sqrt{1-x^2}\right).
\end{equation}
Note that the above integral can be solved analytically when $Q\rightarrow0$ or $Q\rightarrow\infty$. This gives two known asymptotic limits: $f(Q\rightarrow 0)=\frac{\pi Q}{4\sqrt{2}}$ and $f(Q\rightarrow\infty)=2/3$. We find the following expression gives a reasonable fit:
\begin{equation}
    f(Q)=\frac{\frac{\pi}{4\sqrt{2}}Q+1.328Q^2+\frac{2}{3}\times2.79Q^3}{1+2.447Q+3.61Q^2+2.79Q^3},
\end{equation}
with the known asymptotic constraints satisfied. Therefore, the exchange energy can now be approximated by
\begin{equation}
    \epsilon_{x}(r_s,\mu,p)=\frac{-\sqrt{2}}{\pi}\frac{1}{r_s}\left[P_+^3f(Q_+)+P_-^3f(Q_-)\right].
\end{equation}

\section{Static structure factor}\label{appendix:Sk}
\begin{figure*}
\includegraphics[width=1\textwidth]{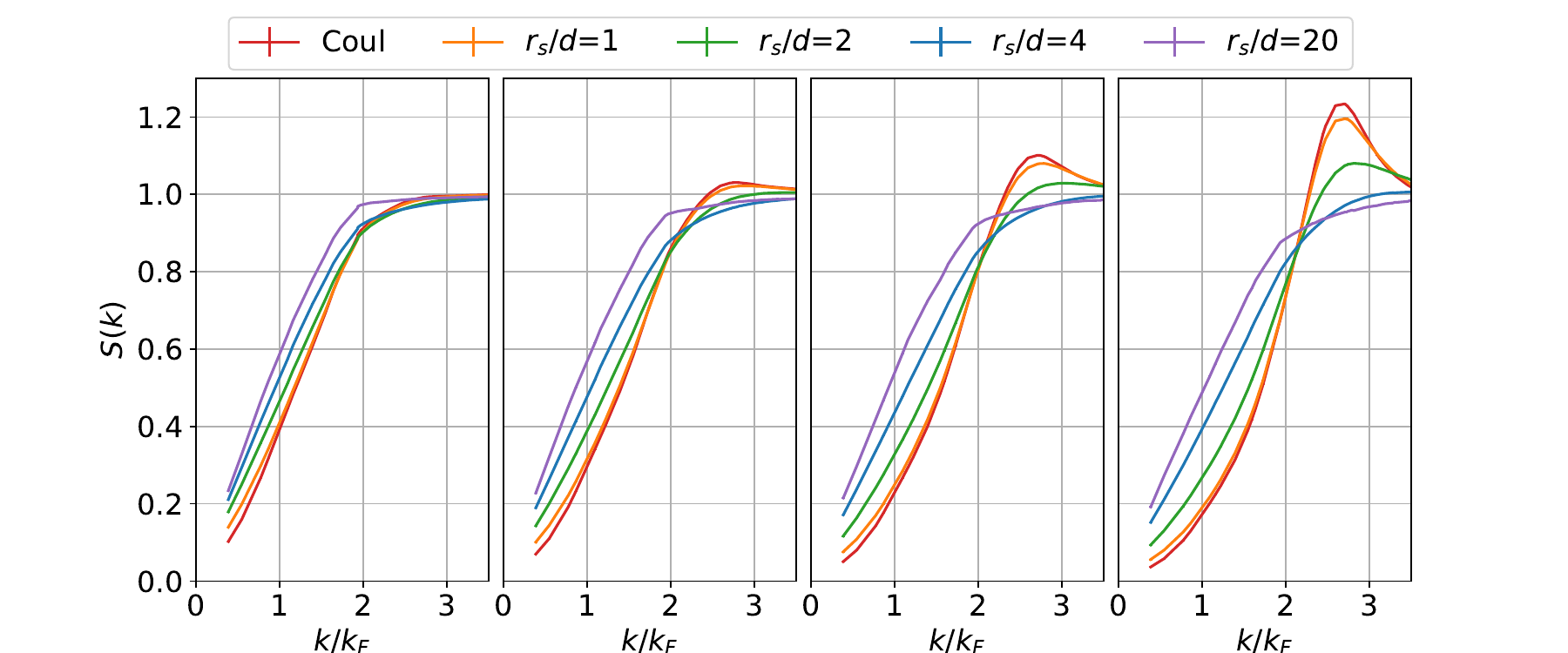}
\caption{The total $S(k)$ obtained under various gate screening in the spin-unpolarized case. The first column is with $r_s=2$, the second column $r_s=5$, the third column $r_s=10$, and the right column $r_s=20$.}
\label{fig:sk0p}
\end{figure*}

\begin{figure*}
\includegraphics[width=1\textwidth]{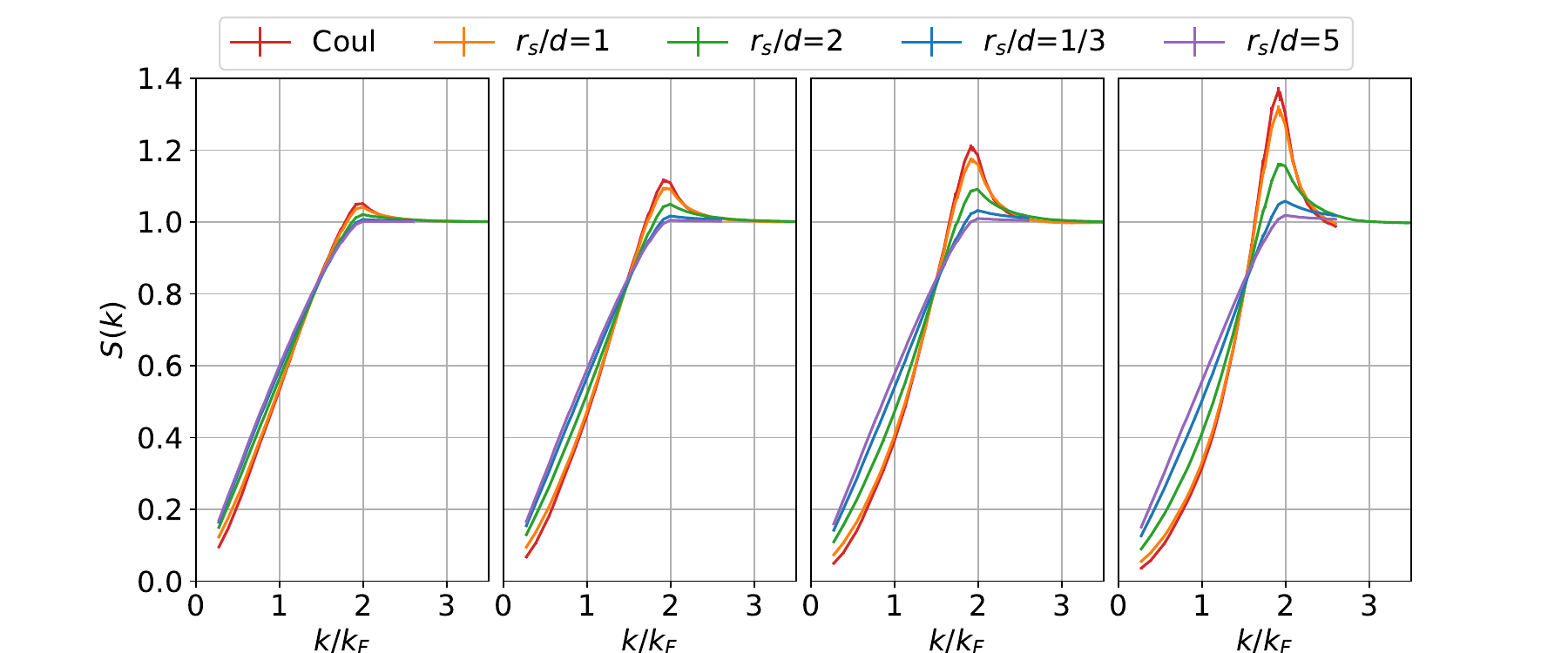}
\caption{The same as Fig.~\ref{fig:sk0p} but for the spin-polarized case.}
\label{fig:sk1p}
\end{figure*}

Line cuts for a range of $r_s$ in the spin-unpolarized case are shown in Fig.~\ref{fig:sk0p}.
The results for the spin-polarized case are shown in \ref{fig:sk1p}.

\end{document}